\documentclass[pra,amsmath,amssymb,showpacs,aps,twocolumn,groupedaddress,nobibnotes
]{revtex4-1}

\usepackage{amsthm}
\usepackage{dsfont}
\usepackage[percent]{overpic}
\usepackage{graphicx}
\usepackage{mwe}
\usepackage{epstopdf}
\usepackage{dcolumn}
\usepackage{bm}
\usepackage{color}
\usepackage[ruled]{algorithm2e}
\usepackage{algorithmic}

\definecolor{myurlcolor}{rgb}{0,0,0.7}
\definecolor{myrefcolor}{rgb}{0.8,0,0}
\usepackage[unicode=true,pdfusetitle, bookmarks=false,bookmarksnumbered=false,
bookmarksopen=false, breaklinks=false,pdfborder={0 0 0},backref=false,
colorlinks=true, linkcolor=myrefcolor,citecolor=myurlcolor,urlcolor=myurlcolor]
{hyperref}

\newenvironment{protocol}[1][htb]
  {% Update algorithm name
   \begin{algorithm}[#1]%
  }{\end{algorithm}}

\newlength\myindent
\newtheorem{prop}{Proposition}\def\PRO{\begin{prop}}\def\ORP{\end{prop}}
\newtheorem{coro}{Corollary}\def\COR{\begin{coro}}\def\ROC{\end{coro}}
\newtheorem{theo}{Theorem}\def \TH{\begin{theo}}\def\HT{\end{theo}}
\def\TH{\begin{theo}}\def\HT{\end{theo}}
\newtheorem{defi}[prop]{Definition}\def\DE{\begin{defi}}\def\ED{\end{defi}}
\newtheorem{lemme}[prop]{Lemma}\def\LE{\begin{lemme}}\def\EL{\end{lemme}}

\usepackage{bbm}

\def\ket#1{\left| #1 \right\rangle}

\def\dm#1{\left|#1 \right\rangle \left\langle #1 \right|}
\newcommand{\ptr}[2]{\text{tr}_{#1}\left[ #2 \right]}

\usepackage[dvipsnames]{xcolor}

\newcommand{\bea}{\begin{eqnarray}}
\newcommand{\eea}{\end{eqnarray}}

\def\k{{\bf k}}

\makeatletter
\newtheorem*{rep@theorem}{\rep@title}
\newcommand{\newreptheorem}[2]{%
\newenvironment{rep#1}[1]{%
 \def\rep@title{#2 \ref{##1}}%
 \begin{rep@theorem}}%
 {\end{rep@theorem}}}
\makeatother
\begin{document}

\title{A quantum network stack and protocols for reliable entanglement-based networks}
\author{A. Pirker}
\author{W.~D\"ur}
\affiliation{Institut f\"ur Theoretische Physik, Universit\"at Innsbruck, Technikerstra{\ss}e 21a, A-6020 Innsbruck, Austria}

\date{\today}

\begin{abstract}
We present a stack model for breaking down the complexity of entanglement-based quantum networks. More specifically, we focus on the structures and architectures of quantum networks and not on concrete physical implementations of network elements. We construct the quantum network stack in a hierarchical manner comprising several layers, similar to the classical network stack, and identify quantum networking devices operating on each of these layers. The layers responsibilities range from establishing point-to-point connectivity, over intra-network graph state generation, to inter-network routing of entanglement. In addition we propose several protocols operating on these layers. In particular, we extend the existing intra-network protocols for generating arbitrary graph states to ensure reliability inside a quantum network, where here reliability refers to the capability to compensate for devices failures. Furthermore, we propose a routing protocol for quantum routers which enables to generate arbitrary graph states across network boundaries. This protocol, in correspondence with classical routing protocols, can compensate dynamically for failures of routers, or even complete networks, by simply re-routing the given entanglement over alternative paths. We also consider how to connect quantum routers in a hierarchical manner to reduce complexity, as well as reliability issues arising in connecting these quantum networking devices. %Finally, we present a detailed example how our quantum network stack can be used to generate graph states across network boundaries.
\end{abstract}

\maketitle

\section{Introduction}\label{sec:intro}

Quantum networks lie at the heart of the success of future quantum information technologies. Possible applications of such networks, over even a global scale quantum internet \cite{KimbleQInet}, are quantum key distribution protocols \cite{ShorQKD,RennerPhd,ZhaoQKD,GottesmanQKD,LoQKD,Wang2015,Chen2016}, quantum conference key agreement \cite{XuQka,SunQka1,SunQka2,Epping2017,Ribeiro18}, secure quantum channels \cite{SecChannelPortmann,SecChannelGarg,SecChannelBroadbent},   clock-synchronization techniques \cite{JoszaClock,Komar14} and distributed quantum computation \cite{CiracDistributed,Caleffi18a,Cacciapuoti18} in general.

In principle there are mainly two approaches to construct quantum networks. On the one hand quantum networks could simply forward quantum information directly, which however needs to be protected against noise and decoherence using quantum error correcting codes \cite{Nielsen}, and repeatedly refreshed at intermediate stations where error correction is performed \cite{Knill96,Zw14,Muralidharan2014,Loock16}. On the other hand, quantum networks may use a property which is only accessible in quantum physics, namely entanglement. Constructing quantum networks by using entanglement has one significant advantage compared to directly motivated approaches: The entanglement topology of a network, which determines in that case also the boundaries and ultimately the structure of a network, is completely independent of the underlying physical channel configuration. In particular, by characterizing quantum networks abstractly in terms of entangled quantum states allows for several interesting features which are not explicitly available in a direct approach, like e.g. creating shortcuts in a network on demand (by introducing an entangled state between parties) \cite{Schoute2016}.

A crucial element to establish long-distance entanglement are quantum repeaters \cite{Br98,DurRepeater,Sa09,Zwerger18,Muralidharan2016,Munro2010,Azuma2015,Pirandola2017}, and multiple proposals for repeater-based networks have been put forward \cite{Epping2016b,Hayashi2007,VanMeter2014book,Pant17a,Das17,Pant17,Munro15,Munro08,Sa11,Guha15,Pirandola16,Meter2013a,Meter2013b,Meter2009,Munro2010,Muralidharan2016} . Most schemes are based on bipartite entanglement, where Bell pairs are generated between nodes of the network. However, a future quantum network shall not be limited to the generation of Bell-pairs only \cite{Meter2011,Pirker18,Cuquet2012,Epping2016a}, because many interesting applications require multipartite entangled quantum states. Therefore, the ultimate goal of quantum networks should be to enable their clients to share arbitrary entangled states to perform distributed quantum computational tasks. An important subclass of multipartite entangled states are so-called graph states \cite{He04}. Many protocols in quantum information theory rely on this class of states.

%One way to construct quantum networks using entanglement is to use bipartite entanglement in form of Bell-pairs. Quantum networks exploiting bipartite entanglement rely on quantum repeater schemes to generate point-to-point entanglement over long distances \cite{Br98,DurRepeater,Sa09,Zwerger18,Muralidharan2016,Munro2010,Azuma2015}.  The fundamental limits of repeater-less communications were investigated in \cite{Pirandola2017}. In the last decade a lot of proposals have been put forward to design architectures and study quantum networks using quantum repeaters \cite{Epping2016b,Hayashi2007,VanMeter2014book,Pant17a,Das17,Pant17,Munro15,Munro08,Sa11,Guha15,Pirandola16,Meter2013a,Meter2013b,Meter2009,Munro2010,Muralidharan2016}

Here we consider entanglement-based quantum networks utilizing multipartite entangled states \cite{Pirker18,Cuquet2012,Epping2016a,Simon2006,Matsuzaki10,Wallnofer16_2D} which are capable to generate arbitrary graph states among clients. In general, we identify three successive phases in entanglement-based quantum networks: dynamic, static, and adaptive. In the dynamic phase, which is the first phase, the quantum network devices utilize the quantum channels to distribute entangled states among each other. Once this phase completes, the quantum network devices share certain entangled quantum states, which results in the static phase. In this phase, the quantum network devices store these entangled states for future requests locally. Finally, in the adaptive phase, the network devices manipulate and adapt the entangled states of the static phase. This might be caused either due to requests of clients in networks, but also due to failures of devices in a quantum network.

We follow the approach of \cite{Pirker18} where a certain network state is stored in the static phase, and client requests to establish certain target (graph) states in the network are fulfilled by processing this network state using only local operations and classical communication (LOCC) in the adaptive phase. This has the advantage that requests can be fulfilled without delay, as the required resource states are pre-generated during the dynamic phase. In contrast, in complete dynamical networks requests are fulfilled by generating the required entanglement over physically available links on demand. This may be rather resource- and time consuming, and involves additional difficulties such as the so-called routing problem \cite{Meter2013b,Schoute2016,Pant17,Gyongyosi2017,Gyongyosi18,Gyongyosi18a,Pirandola16,Das17,Hahn18} where the goal is to determine a way of combining short-distance Bell-pairs to establish a long-distance Bell-pair in the most resource efficient way. %Furthermore, generating several Bell-pairs from a cluster state, which also corresponds to routing, was investigated in \cite{Hahn18}.
In our approach, the problem is split into the generation of a universal network state (which also involves routing, but can be done prior to the request) in the dynamic phase, and its processing in the adaptive phase to establish desired target states. What all approaches have in common are two basic problems:
(i) The complexity of how to organise quantum networking devices in a real network and how to systematically execute tasks in it; (ii) How entanglement can be established efficiently between networking devices in a dynamical manner. Despite the fact that some aspects have been addressed in recent works, it still remains unclear how the different techniques, ranging from the physical channel configuration, over the entanglement structure of a network to routing between quantum networks collaborate to enable for a feasible and tractable quantum network.

%Observe that in general it suffices to construct quantum networks using quantum repeaters generating Bell-pairs, as any graph state can be generated via bipartite entanglement followed by some merging operations. However, two problems appear in that case: First, the clients of the network have to take care of generating the graph state from the Bell-pairs by merging them, thereby having a global knowledge of all quantum networks and introducing noise, and second, the Bell-pairs between spatially separated clients have to be generated by a network of repeaters on demand.

%There have also been some work on quantum networks which generate arbitrary graph states in networks by directly exploiting multipartite entanglement \cite{Pirker18,Cuquet2012,Epping2016a,Simon2006,Matsuzaki10}. For example, in \cite{Pirker18} different architectures for quantum networks relying on multipartite entangled states have recently been introduced. There, quantum networking devices share GHZ or decorated graph states to establish the adjacencies in graph states for their respective clients. Ref. \cite{Cuquet2012} proposes a scheme using GHZ states, but in contrast to \cite{Pirker18}, the clients of the network generate their adjacencies themselves. Finally, \cite{Epping2016a} studies the setting of generating graph states in quantum networks by combining the ideas of quantum repeaters, graph states and quantum error correction.

%We observe that in both cases, bi- and multipartite quantum networks, two problems naturally arise:

Classical computer networks tackle the complexity of transmitting bits between two nodes by breaking down the transmission into several layers of a stack model, the Open Systems Interconnection model (OSI model) \cite{Zimmermann88}. In this model, information passes through seven layers, where each layer has a clear responsibility and adds additional descriptive information to the original message. Networking devices use this prepended information for various tasks. One of these layers is the network layer (layer three), which is responsible for logical addressing and routing in classical networks. Routing protocols for computer networks aim at determining a transmission path from a sender to a receiver by inspecting the descriptive information of the network layer. This task is accomplished by so-called routers, operating on layer three of the OSI model.

The goal of this work is to establish a quantum network stack model from an architectural point of view. We achieve this by abstracting the main concepts which quantum networks necessarily require from their underlying physical implementation details. This provides a clean, and especially technology independent, view on the responsibilities, complexities and tasks arising in quantum networks. Of course, when implementing a quantum network device, one still has to consider how to realize quantum memories, their interfaces to the quantum communication channels, and the implementation of quantum gates. However, in such an abstracted model, implementation details do not affect the concepts residing within the layers of the stack model, since they emerge from a technology independent view on quantum networks. We discuss issues regarding physical implementations in Sec. \ref{sec:stack} in more depth. We believe that a full realization of the quantum network stack as we propose will be challenging in the near-term future. Nevertheless, since the concepts of this work are technology independent, they provide a starting point for building up quantum networks in a well-defined, standardized and technology- and vendor-independent manner.

We introduce a quantum network stack, which breaks down the complexity of entanglement-based quantum networks into several hierarchical layers of a stack model, similar to the OSI model. Each of the layers has a clear responsibility and can therefore be evolved independently in the future. We identify layers for ensuring connectivity at the lowest level (where quantum repeaters operate on), for generating graph states inside a network, but also for enabling inter-network graph state requests. Observe that in such a model, each layer uses its own set of protocols for accomplishing its associated responsibility.

After identifying the quantum network stack, we also present protocols which operate on the proposed layers. We start by proposing a protocol for the static phase in a quantum network for ensuring reliability. Then we discuss protocols tackling problems arising on connecting entanglement-based quantum networks via routers. For that purpose, we introduce the notion of a region, which is essentially a network of routers. The protocol we propose, which we refer to as routing protocol, operates in the adaptive phase and transforms a given entanglement structure between regions to a virtual network state among the requesting networks. This enables our network devices to fulfill arbitrarily distributed graph state requests in a straightforward manner. We also present protocols for the static phase of regions, especially to organize regions in a hierarchical manner and a technique to increase the reliability of regions. Finally, we define the term reachability for quantum networks. In such networks we say that a network (or network device) is quantum mechanically reachable if entanglement to the target can be established or is still present.

The main contributions of our work are as follows:
\begin{itemize}
	\item We introduce a quantum network stack for entanglement-based quantum networks and a classification of quantum network devices in accordance with this stack.
	\item We identify techniques for quantum networks in the adaptive phase to guarantee intra-network reliability, which means that the network devices can deal with the failure of some networking devices without the need to re-establish entanglement.
	\item We present a protocol which is capable of generating arbitrary graph states across network boundaries in the adaptive phase. We also find schemes to tackle the complexity arising in connecting quantum networks as well as reliability issues among regions, both crucial properties for the quantum internet.
	\item We discuss reachability in and between entanglement-based quantum networks.
\end{itemize}

We outline the paper as follows. In Sec. \ref{sec:back} we review necessary background information, some previous works on quantum networks, clarify our assumptions and relate our work to other network and stack models. Next we present our quantum network stack in Sec. \ref{sec:stack}. This includes a discussion of each layer of the quantum network stack as well as auxiliary protocols, but also considerations regarding physical implementations. Then, we discuss reliability for the link layer of our stack in Sec. \ref{sec:layer3:reliable} where we present two techniques for achieving reliability at the link layer as a proof-of-principle and discuss further techniques for ensuring reliability at the link layer. In Sec. \ref{sec:regions} we present protocols for the network layer. Some auxiliary protocols regarding reachability in quantum networks will be discussed in Sec. \ref{sec:aux:reachability}. We provide our conclusions in Sec. \ref{sec:conclusion}. Finally we present concrete, detailed examples of the routing protocol and of how the layers of the stack work together to generate a graph state in Appendix \ref{app:routing} and \ref{app:stack} respectively.

\section{Background}\label{sec:back}

In the following we review relevant background information which we will use in the remainder of this paper. First, we recall how classical computer networks work and describe some classical routing protocols. Then, we introduce Dijkstra's algorithm and Steiner trees on graphs. Next, we define graph states and the effects of applying single qubit unitaries and single qubit measurements on these states. Finally, we review former works on quantum networks which are of relevance for our work.

\subsection{Classical networks}\label{sec:back:classical}

Classical computer networks use the OSI stack model \cite{Zimmermann88}. This model comprises the following seven layers:

\begin{enumerate}
	\item Physical layer: Responsible for the physical transport of bits. This corresponds to different physical transport channels such as wireless technologies (e.g. WIFI, Bluetooth, etc.) or ethernet.
	\item Data link layer: Responsible for managing the access to the physical layer of a network. In particular, protocols at this layer reduce collisions of data transmissions on the physical layer, due to several clients sending data at the same time in a network. This layer introduces hardware addresses, which are unique for every network interface.
	\item Network layer: Responsible for the logical decomposition of a network. Thereby, this layer defines what a logical network is, what its boundaries are and how routing between networks can be done. For example, the IP protocol \cite{ip} introduces the notion of IP addresses on this layer, which identify network devices in a computer network. The network layer adds the IP addresses of the source and destination of a message to the packet, which routers then use to determine a transmission path towards the destination of the packet.
	\item Transport layer: Responsible for controlling the information flow on transporting information between two end nodes of a network. In order to distinguish between different applications/protocols which access the network, the transport layer introduces so-called ports. The destination and source port numbers are added to all packets passing through the transport layer. Furthermore, this layer also provides protocols, like e.g. the Transmission Control Protocol \cite{tcp}, which ensure reliable transmissions between or in networks.
	\item Session layer: Responsible for managing sessions in networks.
	\item Representation layer: Responsible for representing data appropriately.
	\item Application layer: Responsible for displaying the information to the end user (which may also be a computer program). This is the layer which is usually encountered by users of applications, e.g. email clients, web browser, games, etc. They use the lower layers to transport data through a network.
\end{enumerate}

This aims at abstracting implementations details of low levels of the stack w.r.t. higher levels of the stack to provide a simplified view on the network. Such an approach results in several benefits. For example one may evolve only particular layers without the need to consider the complete network implementation and all its details. Another example is that low-level layers are fully independent of high-level layers. More specifically, the network layers does not have to consider any transport control mechanisms which the transport layer poses. This allows developers to change the transport control mechanism without modifying the protocol of the network layer, thereby allowing for flexibility in designing networks.

The third layer, i.e. the network layer, is of special interest in our context, since the information of this layer is used to route packets between two remote points in computer networks. In particular, a client hands over a packet to the network, and the network devices of layer three, which are also called routers, collaborate to determine a path to the recipient. Mainly there are two types of routing protocols: distance-based and link-state protocols. Distance-based protocols, like e.g. Routing Information Protocol (RIP) \cite{RIP}, are not aware of the full network topology, whereas link-state protocols, e.g. Open Shortest Path First (OSPF) \cite{OSPF}, internally keep a graph representation of the current network configuration.  Link-state protocols typically use Dijkstra's algorithm \cite{Dijkstra} to determine the shortest path between two nodes.	

\subsection{Dijkstra's algorithm and Steiner trees}\label{sec:back:algorithms}

In this section we briefly recall two important algorithms in graph theory, which we will use here frequently.

The first algorithm is Dijkstra's algorithm \cite{Dijkstra}. This algorithm is used to determine the shortest path between two vertices $a \in V$ and $b \in V$ of a graph $G=(V,E)$. For that purpose it uses a cost function $f \colon E \to \mathbb{R}_{+}$, which associates with every edge in the graph a certain cost. The algorithm is a greedy algorithm, which evaluates at each step whether a shorter path to a vertex is available. One may generalize Dijkstra's algorithm by not only finding the shortest path to one particular $b \in V$ but for finding a shortest path from $a \in V$ to any vertex $b \in B \subseteq V$. We denote Dijkstra's algorithm throughout this paper by $\mathrm{Dijkstra(a,B)}$ where $a \in V$ and $B \subseteq V$.

The second algorithim we use in this work constructs an approxmiation of a so-called Steiner tree \cite{Steiner} on a graph. A Steiner tree on a graph $G=(V,E)$ is defined for a subset of vertices $S = \lbrace v_1, \ldots, v_k \rbrace$ as a tree which connects the nodes of $S$ with minimal cost. The choice of the cost function $f \colon E \to \mathbb{R}_{+}$ depends on the discussed application. The tree is allowed to also contain vertices which are not an element of $S$. If $S=V$, the algorithm derives a minimal spanning tree.

An algorithm for approximating a Steiner tree for $S$ on $G=(V,E)$ with weighted edges is described in protocol \ref{protocol:steiner}.

\begin{protocol}[h!]
\caption{Steiner($S$, $x$)}
\label{protocol:steiner}
\begin{algorithmic}[1]
    \REQUIRE Set of nodes $S \subseteq V$
    \REQUIRE Starting node $x \in S$
    \STATE $T = (T_S = \emptyset, E_S = \emptyset)$
    \STATE $T_S = \lbrace x \rbrace$
    \WHILE{$T_S \neq S$}
    	\STATE $d_i$ ... distances to $T_S$
    	\FOR{\textbf{each} $x$ in $S \setminus T_S$}
    		\STATE $d_i = \mathrm{Dijkstra(x, T_S)}$
        \ENDFOR
        \STATE $x' = \mathrm{argmin} \, d_i$
        \STATE $T_S = T_S \cup \lbrace x' \rbrace$
        \STATE $E_S = E_S \cup \lbrace \text{Path to } x' \rbrace$
    \ENDWHILE
    \RETURN $T$
\end{algorithmic}
\end{protocol}

The problem of determining a Steiner tree has been shown to be NP-complete in the rectilinear case \cite{SteinerNP}.

\subsection{Graph states and GHZ states \label{sec:back:graph}}

Graph states \cite{He04,Guehne05,He06,Toth06} are an important subclass of multipartite entangled quantum states. These states are associated with a classical graph $G=(V,E)$ where the vertices of $V$ correspond to qubits and the edges in $E$ indicate correlations among the vertices in $V$. In particular, for a graph $G=(V,E)$ (where $|V| = n$) we define the graph state $\ket{G}$ as the common $+1$ eigenstate of the correlation operators
\begin{align}
K_a= X_{a} \bigotimes_{\lbrace a,b \rbrace \in E} Z_{b}
\end{align}
where $X$ and $Z$ denote the Pauli matrices and the subscripts indicate on which qubit the Pauli operator acts on. We call two graph states $\ket{G}$ and $\ket{G'}$ LU-equivalent if there exist unitaries $U_1, \ldots, U_n$ such that $\ket{G} = U_1 \otimes \ldots \otimes U_n \ket{G'}$.

In the following we discuss several important operations on graph states which we will use frequently here \cite{He04,He06}. The first operation is local complementation, which acts just locally on the qubits of the graph state and transforms the graph state according to the following rule: If a local complementation at vertex $a$ is applied, then the subgraph induced by the neighbourhood of $a$ is inverted.

We further also require in this work Pauli measurements of  qubits of a graph state. A measurement in the $Z$ basis effectively removes the measured vertex from the graph, thereby also removing all the incident edges. Depending on the measurement outcome, some local Pauli corrections may have to be applied. A measurement in the $Y$ basis corresponds to the following transformation of the graph state: First a local complementation at the measured qubit is done, followed by removing the corresponding qubit and all its incident edges. Again, depending on the outcome some local Pauli corrections may be required.

Finally we discuss GHZ states. An $n-$qubit GHZ state reads as

\begin{align}
\ket{\mathrm{GHZ}_{n}} = \frac{1}{\sqrt{2}}(\ket{0}^{\otimes n} + \ket{1}^{\otimes n})
\end{align}

This state is local unitary (LU) equivalent to a fully connected graph or a star graph state, see e.g.  \cite{Pirker18}. Here, we usually depict GHZ states using a star graph with a chosen root node. In particular we use the term GHZ state of size $n$, or its state $\ket{\mathrm{GHZ}_n}$, synonymously for the LU equivalent star graph state with $n$ vertices.

An interesting property of GHZ states is that if one combines two GHZ states via a Bell-measurement, then the result is again a GHZ state, up to local Pauli corrections. More specifically, by measuring a qubit of $\ket{\mathrm{GHZ}_{n}}$ and a qubit of $\ket{\mathrm{GHZ}_m}$ with a Bell-measurement leads (up to LU) to the state $\ket{\mathrm{GHZ}_{m+n-2}}$. We will use this property extensively in this work. In principle one can also use a different measurement setup to transform the states $\ket{\mathrm{GHZ}_{n}}$ and $\ket{\mathrm{GHZ}_m}$ to the state $\ket{\mathrm{GHZ}_{m+n-1}}$, see e.g. \cite{Wallnofer16_2D}.

\subsection{Quantum networks}\label{sec:back:networks}

There is extensive work on quantum networks relying on quantum repeaters \cite{Epping2016b,Hayashi2007,VanMeter2014book,Pant17a,Das17,Pant17,Munro15,Munro08,Sa11,Guha15,Pirandola16,Meter2013a,Meter2013b,Meter2009,Munro2010,Muralidharan2016}. Most of these approaches have in common that they assume a network of quantum repeaters sharing Bell-pairs with each other.

However, in general, the goal of a quantum network should be to generate arbitrary states between remote clients rather than solely Bell-pairs. For many applications, it however suffices to be able to generate a specific class of states. Graph states \cite{He04,He06} play an important role in this respect, and the generation of arbitrary graph states among clients has been identified as a desireable goal for quantum networks \cite{Meter2011,Pirker18}. This is what we will require from our network in the following.
For that purpose, several different approaches regarding the entanglement structure may be pursued which include: (i) The usage of a central master node which creates the state locally and teleports it to clients via Bell-pairs shared between the central none and all others; (ii) to establish pairwise entanglement between all of the network nodes first, followed by combining or merging it in an appropriate manner; (iii) generating the target state directly in a distributed manner by using multipartite states.

Approaches (i) and (ii) are far better understood than (iii), due to in-depth knowledge about bipartite entanglement and quantum repeater networks.
Using a central master node which creates the requested graph state and teleports it to clients thereby consuming Bell-pairs basically suffices to generate any arbitrary entangled state between the clients. However, this approach has one significant drawback: If the central master nodes fails, then the whole network is down. This motivates a decentralized approach, leading to quantum repeater networks.

In a network of quantum repeaters sharing bipartite entanglement, depending on the requested target state between the clients of the network, the intermediate quantum repeaters employ entanglement distillation and swapping operations (or other kinds of repeater protocols) to establish the required long-distance Bell-pairs subject to merging to generate the target state. In order to establish this long-distance Bell-pairs, routing in the repeater network needs to be done.

Recently several quantum routing protocols for bipartite quantum repeater networks were presented \cite{Meter2013b,Schoute2016,Caleffi17,Gyongyosi2017,Gyongyosi18,Pirandola16,Das17}. In this context, the goal of a routing protocol is to determine a way of combining short-distance Bell-pairs to establish a long-distance Bell-pair in the most resource efficient way. Ref. \cite{Meter2013b} studies the application of Dijkstra's algorithm to quantum repeater networks, where the cost associated with an edge in the repeater network, i.e. a small-scale Bell-pair, depends on several physically motivated parameters, e.g. Bell-pair generation rate, transmittance, etc. A routing algorithm for ring and sphere type network topologies of repeater networks has been proposed in \cite{Schoute2016}. In \cite{Pant17} a routing protocol for a two-dimensional cluster-type network relying on Bell-pairs was proposed. Another algorithm for optimal routing in an end-to-end setting was subject of study in \cite{Caleffi17}. Routing using an entanglement-gradient in quantum networks was studied in \cite{Gyongyosi2017}. Ref. \cite{Gyongyosi18} constructs a so-called base graph which represents the optimal entanglement structure of a repeater network to determine optimal paths, and a method for adapting this graph, e.g. due to node failures, was studied in \cite{Gyongyosi18a}. In \cite{Pirandola16} lower and upper bounds on the end-to-end capacities in arbitrarily complex quantum networks for single and multipath routing strategies for a single sender and a single receiver, but also for multiple senders and multiple receivers ultimately sharing bipartite states among each other simultaneously were established. Finally, in \cite{Das17} the routing of Bell-pairs in memory-free, two dimensional quantum network was investigated, where intermediate workstations either generate Bell-pairs or perform entanglement swapping, both in configurable directions, thereby achieving routing for establishing Bell-pairs in the network.

Routing protocols using multipartite quantum states received far less attention. Recently it was shown that routing on a cluster state, which is shared among network nodes, using local complementation and measurements in the $X$ basis provides an advantage compared to routing based on Bell-pairs \cite{Hahn18}. Here the main goal was to generate one or several Bell-pairs simultaneously from the cluster state. In addition, they show that by slighting modifying the protocol it can be used to generate GHZ states from the cluster state. An algorithm which closely relates to routing which uses multipartite states was presented in \cite{Dahlberg2017}. In particular, the algorithm of \cite{Dahlberg2017} decides whether a certain stabilizer state (which includes graph states) can be transformed into another stabilizer state by using single qubit Clifford operations, single qubit Pauli measurements and classical communication. In \cite{Dahlberg2018} the complexity of such transformations between graph states were studied and it was shown that this task is in general NP-complete.

\begin{figure}[h!]
\scalebox{3.6}{
\includegraphics{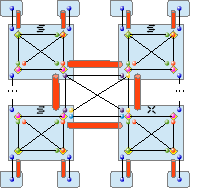}
}
%\flushleft
\caption[h!]{A concrete network example: Three switches (boxes with horizontal arrows) and a router (box with diagonal arrows) connect in a network via GHZ states of decreasing size (black lines indicate entanglement). Internally each of these networking devices use again GHZ states of decreasing size to connect three clients each. In this example, the clients connect via Bell-pairs to the network devices. Observe that the entanglement structure of the network is different from the physical channel configuration (orange tubes). \label{fig:net:example}}
\end{figure}

As discussed in Sec. \ref{sec:intro}, we follow in this work mainly the entanglement-based multipartite approach to quantum networks as presented in \cite{Pirker18}. Therefore, we recall several concepts which were introduced there. Clients are assumed to be of minimal functionality (single qubit unitaries and single qubit measurements), connect to quantum network devices by sharing entanglement with them, for example in form of Bell-pairs. The network devices, i.e. routers and switches, use an internal multipartite quantum state, which we refer to as device state, to connect their clients. Finally, networking devices connect to each other by sharing multipartite entangled quantum states, referred to as the network state. This is illustrated in Fig. \ref{fig:net:example}.

%In the static phase of \cite{Pirker18} is that clients, which are assumed to be of minimal functionality (single qubit unitaries and single qubit measurements), connect to quantum network devices by sharing entanglement with them, for example in form of Bell-pairs. The network devices, i.e. routers and switches, use an internal multipartite quantum state, which we refer to as device state, to connect their clients. Finally, networking devices connect to each other by sharing multipartite entangled quantum states, referred to as the network state. For illustration see Fig. \ref{fig:net:example}.

The aim of the network is to generate arbitrary graph states between its clients on demand. Because the clients have only minimal functionality, the quantum network devices have to carry out the generation of the target state. Since the target state is not known prior to a request, the state of the static phase, i.e. the network and device states, need to be such that any graph state can be generated from them using only LOCC, without generating additional entanglement. All state combinations, i.e. device state and network state, which satisfy this criterion, i.e. that any arbitrary graph state can be generated from them using only LOCC, may serve as a device and network state respectively.

In \cite{Pirker18}, two different types of multipartite states for device and network states were identified: multiple copies of GHZ states (more precisely, the LU equivalent star graph states) of decreasing size, or $m-$partite, fully connected graph state with decoration qubits on each edge (decorated graph state). The number $m$ corresponds to the number of network devices which shall connect. 

Because we will use the GHZ architecture of quantum networks, we clarify this architecture further. In particular, if $m$ network devices connect in this architecture, then the network state corresponds up to local unitaries to

\begin{align}
\bigotimes^{m}_{i=2} \ket{\mathrm{GHZ}_{i}}^{\otimes c_i} \label{eq:back:ghzstate}
\end{align}

where $c_i$ denotes the number of clients which connect to network device $i$. We observe that multiple copies of each GHZ state $\ket{\mathrm{GHZ}_i}$ are mandatory for the network state to enable for arbitrary graph states in the network, see \cite{Pirker18} for a detailed discussion. Furthermore, we do not claim that the state of  (\ref{eq:back:ghzstate}) is optimal, it is just a state from which arbitrary graph states (by including the device states) can be generated in a network using LOCC only. However, depending on the desired target states of the network, only a subset of these states may be required for graph state generation.

We also relate the work of \cite{Pirker18} to the phases of Sec. \ref{sec:intro} as follows: The static phase corresponds to the entanglement structure in the devices and across the  network, i.e. the network states. These states are generated during the dynamic phase, where the networking devices utilize the quantum channels to distribute the required entanglement. One way of doing this is to use the quantum network configuration protocol (QNCP) as discussed in \cite{Pirker18}. In the adaptive phase, the clients request a graph state from the network, and the quantum network devices manipulate the states of the static phase in such a way, that the target state is established between requesting clients. For that purpose, the quantum networking devices apply controlled phase gates (CZ gates), measurements in the $Y$ or $Z$ basis, and Bell-measurements to connect the network and the device state.

Finally we also comment on why following a direct multipartite approach is indeed benefical compared to using Bell-pairs between network devices in our setting. If network devices would share only Bell-pairs among each other, then, depending on the requested graph states, the network devices have to apply more CZ gates, Bell-measurements and single qubit measurements in the $Y$ or $Z$ basis in contrast to a multipartite approach. Observe that if these operations are noisy, this will result in a state of smaller fidelity compared to directly using multipartite quantum states, see \cite{Wallnofer18}. Furthermore, such a direct multipartite approach offers a storage advantage compared to bipartite schemes \cite{Pirker18,Wallnofer18}.

Most works on quantum networks, in the bi- and multipartite case, have in common that there is no clear notion of which quantum task has to be done by which node in the network at which time. In particular, how to organize and classify quantum networking devices depending on their (to-be-defined) capabilities is not fully known yet. However, similar issues arising in quantum repeater networks have e.g. been addressed in \cite{Meter2013a}, which resulted in a stack model for quantum repeater architectures.

\subsection{Assumptions}

For the reminder of this paper we make the following assumptions:

\begin{enumerate}
	\item We restrict our proposal to discrete variable systems in terms of qubits.
	\item Quantum networking devices can apply controlled phase gates, Bell-state measurements, and single qubit Pauli matrices and measurements. 
	\item All quantum operations and measurements are assumed to be noiseless. However, we will discuss possibilities to relax this assumption in Sec. \ref{sec:stack:aux}.
	\item Bell-state measurements are deterministic. We discuss approaches and techniques how to deal with non-deterministic Bell-state measurements in Sec. \ref{sec:stack:con} and Sec. \ref{sec:stack:aux}. 
	\item Quantum network devices have quantum memories. Possibilities how to deal with noise in quantum memory we discuss in Sec. \ref{sec:stack:aux}.
	\item Local quantum states are free, which means that quantum networking devices have the capability to locally create quantum states.	
\end{enumerate}

These assumptions correspond to the points 1-3 and 5-7 of the checklist of \cite{DiVincenzoCheck}. We do not require point 4, since all operations which the network devices apply are Clifford operations, for which several fault-tolerant computing schemes exist \cite{Chamberland18,Chao18}.

\subsection{Relation to other stack and network models}

We relate the main contribution of our work to the closely related works of \cite{Meter2013a} and \cite{Pirker18} as follows. Ref. \cite{Meter2013a} introduces a stack for quantum repeater networks, which establishes single entangled links, i.e. a Bell-pairs, between two nodes of a network. In contrast, the stack we present here in Sec. \ref{sec:stack} aims at constructing arbitrary graph states rather than single links among the clients of the network. This allows the clients of the quantum network to immediately execute more complex protocols, like e.g. conference key agreement or even distributed quantum computation using graph states. The stack of \cite{Meter2013a} may appear inside layer 2 of our quantum network stack, see also Sec. \ref{sec:stack:con}.

Ref. \cite{Pirker18} introduces architectures for single quantum networks, but does not deal with their reliability and issues on connecting different networks to each other. Nevertheless, the GHZ architecture therein, see also Sec. \ref{sec:back:networks}, provides an efficient architecture for quantum networks which reduces storage requirements for network devices by a factor of $2$ compared to direct bipartite architectures using Bell-pairs. However, when connecting several networks the graph state generation process across network boundaries becomes very complex. In order to simplify this process we introduce the concept of region routing in our manuscript. Region routing establishes a virtual network state across requesting network devices, which tremendously simplifies the graph state generation process. More specifically, the output of region routing is a GHZ network state between the requesting network devices, which enables them to directly employ the state linking protocols of layer 3, see Sec. \ref{sec:stack:layer3}, to fulfill the graph state request. We have decided to improve the protocols for GHZ architectures of \cite{Pirker18} as examples of reliability in Sec. \ref{sec:layer3:reliable} and routing in Sec. \ref{sec:regions}.

\section{A quantum network stack}\label{sec:stack}

In classical computer networks, communication in a network follows the OSI layer model, see Sec. \ref{sec:back:classical} for more information. This model vertically breaks down the complexity of networks into several layers. Each of these layers takes data (in form of a packet) from the layer above, and passes it, after optionally prepending the packet by additional descriptive information, to the layer below.

In contrast to the classical network stack where descriptive information is added we assume that qubits of neighbouring layers in the quantum network stack can be accessed and combined.

\begin{figure*}[htpb]
\centering
\scalebox{1.6}{
\includegraphics{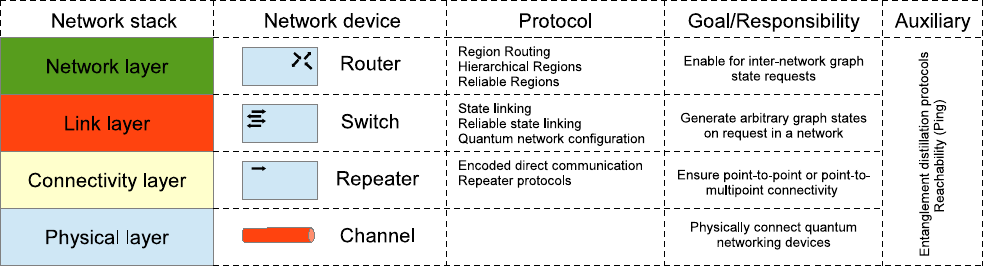}
}
%\flushleft
\caption[h!]{Proposal for a quantum network stack comprising four layers. We identify a physical layer (channel configuration), a conncetivity layer (for ensuring connectivity in terms of high fidelity entangled states between network devices), a link layer (comprosing a single network by sharing a multipartite entangled network state) and a network layer (connecting quantum networks via routers). \label{fig:qstack}}
\end{figure*}

Our proposal for a quantum network stack for quantum networking devices is depicted in Fig. \ref{fig:qstack}.

%\onecolumngrid

%\twocolumngrid
Each layer has a specific goal, meaning that we break down the responsibilities in a quantum network vertically by clearly defining the objectives of a particular layer. One important feature of such a stack is that different layer can be evolved and studied independently. More specifically, changing protocols at higher layers of the stack does not imply changes to the lower layers, i.e. lower layers do not depend on concrete implementations of higher layers. As already shown in Fig. \ref{fig:qstack}, depending on the layer a network device operates on, it has access to more or less layers of the quantum network stack.

The main motivation of following such a stack model in networking are abstractions. 

The key concept which stack models exploit in form of layers is abstraction. In particular, layers introduce boundaries for complexities, as high-level layers do not have to deal with all the details of low-level protocols. For example, a developer of a quantum conference key agreement protocol shall not have to deal with how the network(s) generate the graph state which the key agreement protocol requires. It simply demands the networks to generate such a state between its communication partners, without worrying about all the necessary quantum operations to construct the state, how to establish (possibly long-distance) entanglement or underlying physical implementations (which may also be different between networks). Such an example highlights the necessity (and also power) of abstractions in networks, which we reflect in our quantum network stack. 

For vendors and experimentalists it is of utmost importance to establish such a standardized and technology-independent view on quantum networks, as it provides the community with a set of common protocols and states which quantum network devices can built upon. For example, for quantum network vendors it is necessary to know which operations, protocols and responsibility a quantum network device (need to) has in a network in the end. Without such a common notion or common understanding of a quantum network interoperability of different network devices is not achievable. 

We also elaborate how such a network stack may work in terms of calls and procedures from a programming point of view. In principle, there are two different approaches: synchronous or asynchronous. In a synchronous approach, a high-level layer instructs, or invokes, an operation of a low-level layer and waits until it receives a response. For example, in a synchronous approach, the connectivity layer invokes the operation for establishing short-link entanglement of the physical layer and waits until it responds. In an asynchronous approach, high-level layers invoke operations of low-level layers, but they do not wait for a response of the low-level layer. Instead, the low-level layer notifies the high-level layer by publishing an event that the operation completed. For example, in an asynchronous approach, the connectivity layer invokes the operation for establishing short-link entanglement of the physical layer. However, in the asynchronous case, the connectivity layer does not wait for a response of the physical layer. Instead the physical layer notifies the connectivity layer when the short-link entanglement was established. Such architectural approaches are also referred to as event-driven systems, since the components of a system communicate via events with each other.

%\begin{figure}[h!]
%\scalebox{0.8}{
%\includegraphics{stack_goal.eps}
%}
%%\flushleft
%\caption[h!]{Goals or responsibilities of each layer in the quantum network stack. \label{fig:qstack:goals}}
%\end{figure}

In the following subsections we elaborate on each of these layers in detail. Before doing so we want to emphasize, that if a client of a quantum network has sufficient capabilities, then it may work even on top of the network layer of Fig. \ref{fig:qstack}, which means that it can act as a router or switch in the network. Observe that in such a case, the protocols itself remain unchanged. Furthermore, a client can employ verification techniques and applications to the final states after graph state generation.

We provide a complete example of how the layers of our quantum network stack work together for a particular request in Appendix \ref{app:stack}.

\subsection{Layer 1 -- Physical Layer \label{sec:stack:phys}}

\begin{figure}[h!]
\scalebox{1.5}{
\includegraphics{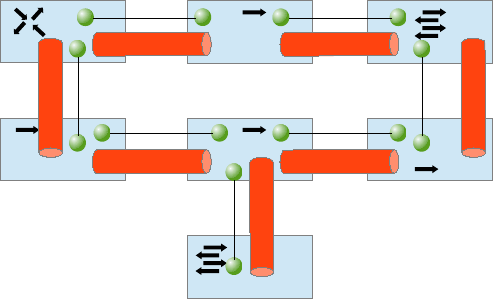}
}
%\flushleft
\caption[h!]{Setting of layer 1 and 2: At layer 1, quantun channels connect quantum networking devices. Neighbouring quantum networking devices operating on layer 2 utilize the quantum channels to create long-distance entangled quantum states. In general, the entanglement structure at layer 2 is independent of the physical channel configuration of layer 1. \label{fig:qstack:12}}
\end{figure}

This layer corresponds to the quantum channels connecting the interacting quantum network devices, for example optical fibres or free space channels. It is responsible for forwarding qubits from one network devices to the other, without applying any error correction or distillation mechanisms. The setting of layer 1 is depicted in Fig. \ref{fig:qstack:12}.

Because this layer deals with the interfacing to the quantum channels of the network, it is also responsible for converting between quantum memory and quantum channel technologies. For example, a quantum network device may store qubits in an ion-trap or NV-center, but for transmission the quantum network device uses an optical setup. Therefore, this layer has to deal with the conversion between these different technologies. Importantly, the layers above, do not have to deal with these interfacing and technology dependent concerns. The physical layer encapsulates these implementation specific interfacing details. 

Quantum network devices will use different communication technologies at the physical layer. Therefore, quantum network converters which convert between the different technologies, or transmission strategies, will be necessary for a full quantum internet. For example, if two networks use different frequencies in an optical setup for transmitting qubits, a converter has to translate between those different transmission strategies.

Finally, we point out that, since this layer is responsible for connecting network devices at the lowest level, the physical layer is responsible for establishing short-link entanglement. For that purpose, several different schemes exist. Some of them do not reveal whether an entangling attempt was successful or not, but some of them, referred to as heralding schemes, provide such a mechanism. Heralding schemes therefore have the advantage that network devices recognize successful entangling attempts, which enables the physical layer to repeat entangling attempts until success. Regarding the distribution protocol for short-entanglement, quantum network devices may use any protocol which enables for heralding like e.g. Meet-in-the-middle or Sender-Receiver of \cite{Jones2016}, parts of the multiplexing protocol of \cite{Munro2010} or the protocols of \cite{VanDam2017} in case of optical transmission setups. Such protocols can run until entanglement attempts complete successfully, which the physical layer then reports to the connectivity for further processing.

This loose coupling between layers perfectly fits into stack models, in which high-level layers do not care about how lower layers fulfill their tasks, they only care about that tasks complete (by whatever means). In particular, the connectivity layer (which will be responsible for generating long-distance entanglement) assumes that short-link entanglement was generated, but the connectivity layer shall not be concerned about how it was generated.

\subsection{Layer 2 -- Connectivity Layer \label{sec:stack:con}}

This layer tackles errors due to imperfections in the quantum channels of layer 1. On this layer, the techniques for establishing long-distance quantum communications reside. In particular, concrete technologies include quantum repeaters, bi- \cite{Br98,DurRepeater,Sa09,Zwerger18,Azuma2015,Epping2016b,Hayashi2007,VanMeter2014book,Pant17a,Das17,Pant17,Munro15,Munro08,Sa11,Guha15,Pirandola16,Meter2013a,Meter2013b,Meter2009,Munro2010,Muralidharan2016}  or multipartite \cite{Wallnofer16_2D,Wallnofer18}, but also the direct transmission of encoded quantum states \cite{Knill96,Zw14,Muralidharan2014,Loock16} or percolation approaches \cite{AcinNetwork,Rudolph17} which generate entanglement structures in a noisy quantum network of networking devices by applying techniques from percolation theory. The main purpose of devices operating on this layer is to enable point-to-point or point-to-multipoint long-distance connectivity without any notion of requests. This functionality is crucial for the dynamic phase of a quantum network, where network devices have to (re-)establish long-distance entanglement, i.e. network states, required in the static phase. Observe that, if quantum repeaters are in use, the entanglement structure in terms of Bell-pairs can be independent (or different) from the configuration of quantum channels at layer 1.

The notion of success here depends on the protocol the connectivity layer employs. For example, in case of quantum repeaters which use recurrence-type entanglement distillation, the long-distance entanglement attempt can fail if the distillation step of the recurrence-type protocol fails. However, the quantum repeaters detect such a fail due to classical outcomes during protocol execution, which enables them to re-iterate the distillation step until the complete distillation protocol successfully completes. Furthermore, if Bell-state measurements are non-deterministic, the repeater protocol may also fail on performing entanglement swapping. However, several different protocols exist to deal with non-deterministic Bell-state measurements and for recognizing successful entanglement attempts for quantum-optical implementations \cite{VanEnk97,VanEnk98}.

Higher layers, i.e. devices operating on a higher layer like e.g. switches or routers, utilize this layer to establish multipartite entangled quantum states within a network or between several independent networks, see Fig. \ref{fig:qstack:12}.

The layer above, i.e. the link layer, is independent of the protocol which this layer uses. It simply instructs this layer to perform certain tasks within the network, like e.g. establishing a long-distance Bell-pair or a GHZ state between other high-level networking devices. Such instructions may also involve several devices of this layer across the network. Such an abstraction enables quantum network administrators to easily change protocols. For example, the link layer is not aware of whether the network devices use quantum repeaters, send encoded states across the quantum channels or rely on techniques from entanglement percolation to generate entanglement. 

Since the layers above (link and network layer) are completely decoupled from the connectivity layer, devices at layer 2 can apply enhanced techniques, like e.g. finding optimal paths in quantum repeater networks (see Sec. \ref{sec:back:networks}) by routing, without affecting upper layers of our stack.

\subsection{Layer 3 -- Link layer}\label{sec:stack:layer3}

\begin{figure}[h!]
\scalebox{1.5}{
\includegraphics{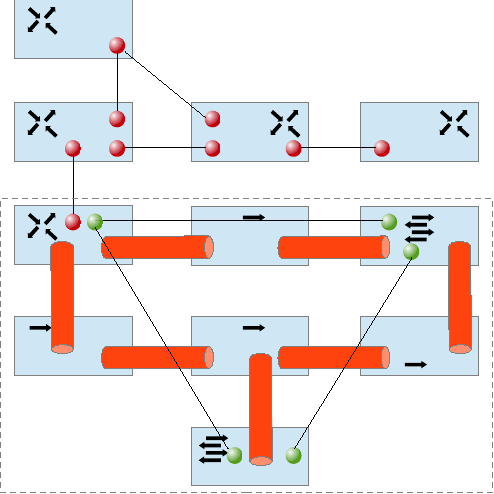}
}
%\flushleft
\caption[h!]{Setting of layer 3 and 4: The network devices of layer 3 request and combine the entangled states from layer 2 to create the network state (green nodes). Depending on client requests, this network state is consumed during graph state generation. On layer 4, quantum routers connect quantum networks via multipartite entangled quantum states (red nodes). In this figure, the quantum routers connect via GHZ network states. \label{fig:qstack:34}}
\end{figure}

The link layer defines the boundaries of a quantum network in terms of an entangled, distributed, multipartite network state which the networking devices of a quantum network share in the static phase. This layer utilizes the connectivity layer to establish the entangled network state during the dynamic phase, which therefore enables also for long-distance quantum networks. Once the dynamic phase completes, the link layer devices (switches) share multipartite entangled states which comprise the network state, see Fig. \ref{fig:qstack:34}. Thereby we end up in the static phase. We observe that the entanglement structure can be completely different from the underlying configuration of quantum channels and devices or protocols operating at layer 2. Concrete instances of networks may connect in the static phase via e.g. GHZ states or decorated graph states (see also Sec. \ref{sec:back:networks}), as proposed in \cite{Pirker18}.

The link layer orchestrates and coordinates the process of generating the network state, utilizing the connectivity layer, in the dynamic phase and is responsible for generating arbitrary graph states between clients of the network during the adaptive phase via a so-called linking protocol, see Fig. \ref{fig:qstack}. The linking protocol is responsible for transforming the entangled network state and device-internal states using only LOCC to the requested graph state, which consumes the entanglement of the network state. This linking protocols depend on the network state. For example, for GHZ or decorated architectures, devices at this layer may invoke the linking protocols of \cite{Pirker18} to create the requested graph state.

Furthermore, the link layer also has the capability to invoke entanglement distillation protocols for two-colorable graph states \cite{Du03,Du05,Duer07,Chen07} which ensure that the network state has a sufficiently high fidelity. In addition it uses the auxiliary protocols for entanglement swapping and merging states, which we discuss in Sec. \ref{sec:stack:aux}.

\subsection{Layer 4 -- Network layer}\label{sec:stack:layer4}

The network layer is responsible for generating and manipulating inter-network entanglement to enable graph state requests spanning several different quantum networks. The network devices operating on this layer are quantum routers. They are connected with each other via multipartite entangled quantum states in so-called regions in the static phase, similar to quantum networks at the link layer. The corresponding states depend on the protocol of this layer. The overall setting is illustrated in Fig. \ref{fig:qstack:34}.

%\begin{figure}[h!]
%\scalebox{1.5}{
%\includegraphics{layer_4.eps}
%}
%%\flushleft
%\caption[h!]{Setting of layer 4: Quantum routers connect quantum networks via multipartite entangled quantum states. In this figure, the quantum routers connect via GHZ network states. \label{fig:qstack:4}}
%\end{figure}

Regions connect in the same fashion as quantum networks do. More specifically, quantum routers in the same region share a multipartite entangled quantum state with each other, like e.g. a GHZ network state, but in contrast to a switch, a quantum router may be part of several regions at the same time. Furthermore, a router may also be part of a quantum network of the link layer, thereby providing an entry point to that quantum network from the viewpoint of other networks. We outline available operations and protocols below, and discuss them in more depth in Sec. \ref{sec:regions} for regions connecting via GHZ states. Observe that lower layers, like e.g. link or connectivity layer, are independent of the protocols and considerations of the network layer.

The network layer is responsible for enabling for graph state requests across network boundaries in the adaptive phase. Therefore, some sort of routing between different quantum networks needs to be done. More precisely, to enable for graph state requests across networks, a quantum routing protocol should establish a "virtual network state" between the quantum networks which are part of a graph state request. The topology of the virtual network state will depend on the routing protocol. Note however that this only involves local manipulation of entangled states that are already present in the network from the static phase, and no generation of additional entanglement is required. Hence these requests can typically be fulfilled fast.

Once routing finishes, the routers use this "virtual network state" of the network layer to establish a network state between the requesting network devices inside their respective networks. Routers achieve this by combining the virtual network state of layer 4 with the inner network state of layer 3 by local operations, for example in terms of Bell-measurements for GHZ states or controlled phase gates and measurements in the $Y$ basis for decorated architecture. However, the output of the routing protocol creates a full network state among the requesting network devices. %For the requesting network devices the routing process is completely transparent.
It might be necessary to transform the output state of routing into an appropriate form to combine it with the intra-network states. We note that the link layer itself is not involved in this routing process.

We discuss a routing protocol for regions using GHZ network states in Sec. \ref{sec:regions:routing} and approaches for simplifying the complexity in regions and introducing reliability for connecting routers using GHZ network states in regions in Sec. \ref{sec:regions:hierarchical} and Sec. \ref{sec:regions:reliable} respectively.

Finally, the network layer can also invoke entanglement distillation protocols for two-colorable graph states, and the techniques for entanglement swapping and merging from the set of auxiliary protocols at all layers, see Sec. \ref{sec:stack:aux}.

\subsection{Auxiliary protocols}\label{sec:stack:aux}

As illustrated in Fig. \ref{fig:qstack}, each layer has access to some auxiliary protocols. The network devices use these protocols to e.g. to (i) generate high-fidelity entangled quantum states, (ii) check whether network devices operating at the same layer are still reachable, (iii) perform entanglement swapping or merging, (iv) employ techniques for error correction, or (v) classically monitor the status of the network using the techniques of (i)--(iv). Depending on the layer a quantum network device operates on it will use different subsets of the aforementioned protocol types. 

The protocols of (i) are entanglement distillation protocols. The layers use these protocols to generate high-fidelity entangled quantum states across the network by transforming several noisy input states to fewer, but more entangled copies. We can associate to each layer one class of entanglement distillation protocols. For example, layer 2 may use entanglement distillation protocols for Bell-pairs \cite{BennettRecurrence,Deutsch}, whereas layer 3 and layer 4 need access to the entanglement distillation protocols for two-colorable graph states \cite{Du03,Du05,Duer07} or CSS states \cite{Chen07}.

The protocols of (ii) address reachability in quantum networks. In entanglement-based quantum networks, there are different forms of reachability. At a basic level, this is about the (classical) reachability of the corresponding network device. However, this is not sufficient as also the presence of the required entangled states needs to be ensured. We discuss this issue in detail in Sec. \ref{sec:aux:reachability}.

Entanglement swapping and merging, i.e. protocol type (iii), are operations that are crucial for repeater architectures, but also for the modification of entangled states on the link and network layer. Entanglement swapping corresponds to a Bell-measurement which is applied to one qubit of two Bell-states each plus the classical communication of the measurement outcome. Such a measurement establishes again a Bell-pair between outer nodes and is usually used to generate long-distance Bell-pairs. But it may also be used to combine two GHZ states into a single larger GHZ state. In contrast, merging connects two graph states into a single graph state in a well-defined, protocol-dependent manner. This technique emerges especially at the link layer, at which network devices execute linking protocols to generate graph states which clients request. Such protocols include controlled phase gates, as well as single qubit measurements. Observe that the realization of such operations (Bell-state measurements, controlled phase gates, single qubit measurements) depend on the physical implementation technology. Therefore, these types of auxiliary protocols may include fault-tolerant quantum computational elements to deal with noise. Furthermore, the network devices may employ different strategies for dealing with non-deterministic gates and measurements, like e.g. \cite{Browne05} for optical implementations or \cite{Kieling07,Zaidi15}.

The techniques of (iv) correspond to quantum error correcting codes, which networking devices may use to tackle channel noise and loss which occurs during the dynamic phase, but also on storing qubits which are part of larger entangled states for a longer time in quantum memory in the static phase.

The techniques of (v) monitor the health-status of the stack layers, also across device boundaries. For that purpose we use the protocols of (i) to ensure that the fidelity of quantum states is sufficiently high. Monitoring the fidelity of quantum states is in general difficult, however, devices may employ parameter estimation techniques to statistically infer the fidelity of the entire ensemble of quantum states by employing measurements to a sub-ensemble. Furthermore, the protocols of (ii) may be used on a regular basis to decide whether network devices are reachable. If devices do not answer to these reachability requests, the remaining network devices conclude that they are no longer part of the network, thereby invoking recovery mechanisms, see e.g. Sec. \ref{sec:linking:shield}.

Additional auxiliary protocols may be added on demand. We emphasize the importance of these kind of protocols in a network stack, since monitoring the health status of a network as well as the recovery from failures, are indispensable mechanism to operate a full functioning network.  

\section{Layer 3 -- Reliable state linking}\label{sec:layer3:reliable}

In this section we discuss how to achieve reliability at the link layer using multipartite entanglement, which is of high importance for the static phase of a quantum network. The term reliability means in our case that parts of the entanglement structure in a quantum network remain intact, i.e. usable for other devices, if one network device disconnects without performing any further operation. %Recall that the link layer is responsible for generating intra-network graph states on request.

Before discussing the reliability of multipartite networks, we review its issues arising in bipartite networks using Bell-pairs. In this case, reliability highly depends on the topology of the distributed Bell-pairs.

For example, consider a quantum network with a central master node sharing Bell-states with all clients. Clearly, if this central master node disappears, all Bell-pairs are lost, and hence, no further communication is possible \footnote{Such an effect may be tackled by introducing several master nodes, which implies that large quantum memories will be required at these master nodes}. In a fully bipartite approach, where all quantum network devices connect to each other via Bell-states in a decentralized manner, and clients only connect to these network devices, we note that the problem of reliability disappears. The failure of a node only affects the entangled pairs the node is part of, all other Bell pairs remain undisturbed.

Here we consider the GHZ architecture within networks for our reliability protocols in quantum networks, see \cite{Pirker18}. Recall that in this architecture network devices which reside within the same network share multiple copies of GHZ states of decreasing size. The network state connecting $m$ devices is, up to several copies of the states, local unitary equivalent to the state 
\begin{align}
\bigotimes^{m}_{i=2} \ket{\mathrm{GHZ}_{i}}. \label{eq:ghznet}
\end{align}
In particular, the network state corresponds to the star graph states which one obtains by transforming each GHZ state $\ket{\mathrm{GHZ}_{i}}$ of (\ref{eq:ghznet}) to a star graph state of size $i$ via local unitaries. Due to this local unitary equivalence, see also Sec. \ref{sec:back:graph}, we often use in the remainder of this work the term GHZ state to refer to the corresponding star graph state.

If a network device leaves the network, it measures all of its leaf qubits of the GHZ states of (\ref{eq:ghznet}) in the $Z$ basis and the root qubits of its associated GHZ states of (\ref{eq:ghznet}) in the $X$ basis. These measurements simply reduce the sizes of all GHZ state connecting the devices, thereby preserving entanglement in the network.

\begin{figure}[h!]
\scalebox{3.5}{
\includegraphics{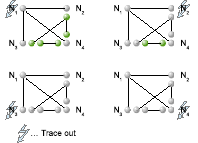}
}
%\flushleft
\caption[h!]{GHZ states are very fragile: If one of the network devices disappears, then at least one GHZ state is completely lost (grey vertices -- entanglement lost), as disconnecting corresponds to tracing out a qubit from a GHZ state which results in a separable state. The green vertices are not affected by the disconnect. \label{fig:ghz:fail}}
\end{figure}

We now illustrate why it may be a problem to directly use star graph states without any further modification as network state. For that purpose, consider the GHZ state $\ket{\mathrm{GHZ}_i}$, and suppose one network device disconnects from the network without performing the protocol for leaving the network on his qubits of the network state. Such a disconnect corresponds to tracing out all the qubits of that particular network device. The state after tracing out any qubit of one GHZ state of (\ref{eq:ghznet}) results in

\begin{align}
\ptr{j}{\dm{\mathrm{GHZ}_i}} = \frac{1}{2}(\dm{0}^{\otimes (i-1)} + \dm{1}^{\otimes (i-1)}),
\end{align}
which is a separable state. Therefore, losing one qubit due to a disconnect will destroy the entanglement between all other network devices which are part of that GHZ state. The situation is shown in Fig. \ref{fig:ghz:fail}. As depicted in Fig. \ref{fig:ghz:fail}, depending on which network device disconnects, we may even loose all network states. In particular, if one of the network devices which connect via $\ket{\mathrm{GHZ}_{2}}$ (i.e. $N_3$ or $N_4$) disappears, all network states will be lost, since these network devices store one leaf of each GHZ state of (\ref{eq:ghznet}).

Because we cannot predict which network device will fail or disconnect, we have to find solutions which are able to deal with the disconnect of any of the network devices such that the functionality for the remaining system is preserved. Nevertheless, we find that schemes using multipartite entanglement are still more beneficial in terms of storage size compared to a full bipartite approach. In the following we discuss two protocols as a proof-of-principle which tackle the effect of failing network devices.

\subsection{Reliable state linking -- Symmetrization \label{sec:linking:sym}}

In general, several copies of each GHZ states in (\ref{eq:ghznet}), which comprises the network state, are mandatory to enable for arbitrary graph state requests in a network.

\begin{figure*}[htpb]
\scalebox{5}{
\includegraphics{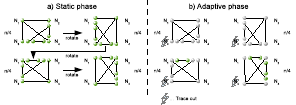}
}
%\flushleft
\caption[h!]{a) Static phase: The first solution to tackle device failures  is to rotate the full network state, i.e. cyclically shift the parties. For $m$ network devices cyclically shifting the network state $m$ times results in $m$ different configurations. More specifically, we split the $n$ copies of (\ref{eq:ghznet}) into $m$ configurations, where each configuration is obtained by cyclically shifting the root-leaf assignment in (\ref{eq:ghznet}). A further extension might be a full symmetrization among the network devices. b) Adaptive phase: If one network device disconnects, then there exists at least one configuration which ensures full connectivity among the remaining devices $\longrightarrow$ at least $n/m$ copies of the network state of (\ref{eq:ghznet}) remain intact. But there exist also other configurations remaining partially intact.  \label{fig:ghz:sym}}
\end{figure*}

The first solution we propose is to symmetrize the network state. In particular, we circularly shift the parties of the network w.r.t. to their assignment to leafs and roots of the GHZ states of Eq. (\ref{eq:ghznet}). The situation is summarized in Fig. \ref{fig:ghz:sym}.

Observe that by symmetrizing the root-leaf assignment in the state of (\ref{eq:ghznet}), each network device stores several roots of each GHZ state $\ket{\mathrm{GHZ}_{i}}$ where $2 \leq i \leq m$ and $m$ denotes the number of devices in the network. We call the state obtained after one cyclic shifting step a configuration.

The crucial observation is that if one network device disconnects, there exists one configuration for which the disconnecting device holds the root of the largest GHZ state $\ket{\mathrm{GHZ}_{m}}$ of (\ref{eq:ghznet}). All other network devices connect in this configuration via the states $\ket{\mathrm{GHZ}_{i}}$ where $2 \leq i \leq m-1$. Therefore, the disconnect of this network device only destroys the largest GHZ state in that configuration whereas all other states in this configuration remain intact. The situation is summarized in Fig. \ref{fig:ghz:sym}. 

Furthermore, as we discuss in Fig. \ref{fig:ghz:sym}, we propose to distribute $n$ copies of the state of  (\ref{eq:ghznet}). Because the number of network devices, i.e. $m$, is constant (unless a network device leaves the network, which we assume to happen rarely as also in classical networks), at least $n/m$ copies of the state in (\ref{eq:ghznet}) remain intact. Therefore, by increasing the number of copies $n$, the network administrator is able to attain higher reliability for the quantum network. For example, by letting $n=2m$ the network administrator ensures that in the case that one of the network devices fails, at least $2$ full copies of the network state of (\ref{eq:ghznet}) remain intact for further processing. We also observe that in case of symmetrization the protocol for leaving the network for a network device does not change.

Several variants of such a symmetrization approach are possible. For example, it is worth considering instead of symmetrizing GHZ states of different sizes, as we propose above, to symmetrize GHZ states of constant size in a uniform way across the network devices. While such an approach is beneficial in terms of reliability, it introduces an additional overhead in terms of resources, i.e. qubits which the network devices have to store, to ensure the goal of  a quantum network which is the generation of arbitrary graph states between clients. Another variant, in order to tackle arbitrarily losses and including failures of multiple network devices, is to use a full symmetrization according to all possible permutations of network devices with GHZ states of decreasing size. 

\subsection{Reliable state linking -- Shielding \label{sec:linking:shield}}

\begin{figure*}[htpb]
\scalebox{5}{
\includegraphics{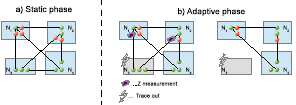}
}
%\flushleft
\caption[h!]{a) Static phase: The second solution to achieve reliability in a quantum network using entanglement in form of GHZ states is to introduce shield qubits. We place on each edge of every GHZ state (represented by a star graph) one qubit and consider the resulting graph state. Only the Bell pair needs no decoration, as entanglement is gone in case of failure of either of the two nodes. b) Adaptive phase: Because tracing out a qubit commutes with $Z$ measurements on other qubits, the remaining devices just have to measure the shield qubits to the disconnected device in the $Z$ basis. The remaining devices will still have a full network state.\label{fig:ghz:shield}}
\end{figure*}

The second solution to ensure reliability in a quantum network is to introduce shielding qubits to the star graph states comprising the network state. In the following, when referring to a GHZ state $\ket{\mathrm{GHZ}_i}$ we mean the corresponding star graph state of size $i$. To achieve reliability, we place on each edge of the GHZ states $\ket{\mathrm{GHZ}_i}$ in (\ref{eq:ghznet}) one additional qubit (which we call the shielding qubit of that edge in the GHZ state), except the Bell-pair, of the network state. We consider the graph state corresponding to this decorated graph, where we use the star graph to represent the initial GHZ state. This shielding qubit belongs to the network device which holds the root of the respective GHZ state, see Fig. \ref{fig:ghz:shield}. In terms of stabilizers we uniquely describe the corresponding graph state resulting of shielding the GHZ states $\ket{\mathrm{GHZ}_i}$ for $2 \leq i \leq m$, where $m$ denotes the number of network devices, as the eigenstate of the family of operators

\begin{align}
K_j=X_j \bigotimes_{k \in N_j} Z_k  \label{eq:ghz:shield:stab}
\end{align}

where $j$ denotes the vertices of the graph state in the static phase of Fig. \ref{fig:ghz:shield}, and $N_j$ the neighbourhood of vertex $j$. 

To create shielded GHZ state $i$, network device $i$ locally prepares a star graph state of size $i$, as well as long-distance Bell-pairs to the network devices $1 \leq j \leq i-1$ using the connectivity layer. Finally, network device $i$ merges its local star graph state with the Bell-pairs which results in the state of (\ref{eq:ghz:shield:stab}). Alternatively, if the connectivity layer uses the transmission of encoded states, network device $i$ prepares the shielded GHZ state locally and transmits the leaf qubits as encoded states to the network devices $1 \leq j \leq i-1$.

During network operation, the link layer detects network device failures by using classical ping messages, see monitoring protocol in Sec. \ref{sec:stack:aux}. For example, if a network device is not responding to the ping requests, then all other network devices assume that the unreachable network device has disconnected. The crucial observation for the shielded GHZ states of Fig. \ref{fig:ghz:shield} is that, if a network device disconnects, which corresponds to a trace out of its qubits, the remaining network devices preserve their states by measuring all the shielding qubits to the disconnecting network device in the $Z$ basis. This can easily be seen since the trace out operation commutes with the $Z$ measurement of neighbouring qubits, which effectively decouples the part of the shielded GHZ network states corresponding to the disconnecting network device, see Fig. \ref{fig:ghz:shield}. Because we assume that operations are deterministic, the recovery operation from an unexpected disconnect is also deterministic.

%\begin{figure}[h!]
%\scalebox{2.8}{
%\includegraphics{ghz_net_shield_ex.eps}
%}
%%\flushleft
%\caption[h!]{Because tracing out a qubit commutes with $Z$ measurements on other qubits, the remaining devices just have to measure the shield qubits to the disconnected device in the $Z$ basis. The remaining devices will still have a full network state. \label{fig:ghz:shield:ex}}
%\end{figure}

However, if no error occurs, the networking devices can reduce the shielded GHZ states to a GHZ state by measuring the shielding qubits in the $Y$ basis, which establishes wires to other network devices, see e.g. \cite{Pirker18}. Observe that depending on the measurement outcomes, some Pauli corrections may be necessary.

We compare the number of qubits necessary in this shielded GHZ approach to a full bipartite solution, see beginning of Sec. \ref{sec:layer3:reliable}, solely using Bell-pairs, because this scheme automatically ensures reliability in quantum networks.

In \cite{Pirker18} it was shown that the number of qubits of a GHZ network state for a network of $m$ devices connecting $c_1, \ldots, c_m$ clients respectively after expanding the network state to all connected clients is

\begin{align}
M_{M} = \sum\limits^{m}_{i=2} \left[ c_i \left(1 + \sum\limits^{i-1}_{k=1} c_k \right) \right]. \label{eq:ghzqubits}
\end{align}

We explain (\ref{eq:ghzqubits}) as follows: Network device $i$ connects to the network devices $1, \ldots, i-1$ via $c_i$ copies of the GHZ state $\ket{\mathrm{GHZ}_{i}}$. Each copy of that GHZ state corresponds to the adjacency of one client of network device $i$ to the $c_{1}, \ldots, c_{i-1}$ clients located at the network devices $1, \ldots, i-1$. To take into account for all this adjacencies, the network devices $1, \ldots, i-1$ expand each of the $c_i$ copies of the GHZ state $\ket{\mathrm{GHZ}_{i}}$ to $\ket{\mathrm{GHZ}_{1 + \sum^{i-1}_{k=1} c_k}}$ via Bell-measurements. We refer to this state also as expanded network state.

Recall that we decorate each edge of the GHZ network state $\ket{\mathrm{GHZ}_{m}}, \ldots, \ket{\mathrm{GHZ}_3}$ once, and that device $i$ has $c_i$ copies of the state $\ket{\mathrm{GHZ}_i}$. Therefore, the total number of qubits which have to be stored including the shielding qubits is

\begin{align}
M_{S} = \sum\limits^{m}_{i=2} \left[ c_i \left(1 + \sum\limits^{i-1}_{k=1} c_k \right) \right] + \sum^{m}_{i=3} c_i (i-1).
\end{align}

For the number of qubits necessary in following a direct bipartite approach one finds (see \cite{Pirker18}) that

\begin{align}
M_{B} = 2 \sum\limits^{m-1}_{i=1} c_i \sum\limits^{m}_{j=i+1} c_j
\end{align}

qubits are required in total. We compare the number of qubits of the shielded GHZ network state and the bipartite approach for various scenarios in Table \ref{tab:shield}.

\begin{table}[h!]
\begin{tabular}{l|l|c|c|c}
 $c$ & $m$ & $M_B$ & $M_S$ & $M_M$ \\
\hline & $m=5$ & $180$ & $129$ & $102$ \\ %$285$ \\
 $c=3$ & $m=10$ & $810$ & $564$ & $432$ \\ %$1245$ \\
 & $m=15$ & $1890$ & $1299$ & $987$ \\ %$2880$ \\
\hline & $m=5$ & $500$ & $315$ & $270$ \\ %$775$ \\
 $c=5$ & $m=10$ & $2250$ & $1390$ & $1170$ \\ %$3425$ \\
 & $m=15$ & $5250$ & $3215$ & $2695$ \\ %$7950$ \\
 \hline & $m=5$ & $980$ & $581$ & $518$ \\ %$1505$ \\
 $c=7$ & $m=10$ & $4410$ & $2576$ & $2268$ \\ %$6685$ \\
 & $m=15$ & $10290$ & $5971$ & $5243$ %$15540$
\end{tabular}
\caption{\label{tab:shield} In this table we compare the number of qubits which have to be stored in a direct bipartite approach solely using Bell-pairs to the shielded GHZ network state, and the GHZ network state without shielding with different number of clients $c_i = c$ and $m$ devices.}
\end{table}

From Table \ref{tab:shield} we find that even though it seems at first glance that shielding the GHZ network state will introduce a large overhead, it still results in better performance in terms of qubits to be stored compared to a direct bipartite approach. The reason for this is that we only place qubits on the edges of the network state before expansion, and not for the expanded network state, see paragraph below Eq. (\ref{eq:ghzqubits}).

In contrast to symmetrization, shielding requires shielded GHZ states instead of GHZ states as a network state. These shielded GHZ states impose an additional overhead in terms of quantum memory compared to the symmetrization technique of Sec. \ref{sec:linking:sym}. Nevertheless, at the same time, shielding is more effective in case of device failures, as all states of the network remain intact after a device failure.

Finally, we note that the protocol for leaving a network changes slightly in case of reliable state linking with shielding. In particular, if a network device leaves the network, it first measures all of its shielding qubits in the $Y$. Then it execute the protocol for leaving a GHZ network.

\subsection{Further considerations \label{sec:linking:further}}

The two protocols we presented in Sec. \ref{sec:linking:sym} and Sec. \ref{sec:linking:shield} are meant as a proof-of-principle how to achieve reliability in a quantum network using multipartite quantum states. However, there might exist several other approaches which enable for reliability in quantum networks. For example, the network architecture using decorated graph states in \cite{Pirker18} is reliable in the sense we discuss here. In addition one may also consider to employ networking coding techniques, like e.g. in \cite{Epping2016b} to achieve reliability in a quantum network. Further extensions to quantum error correction codes or other approaches may also be viable, which we leave for future works.

\section{Layer 4 -- Region routing, Hierarchical regions and Reliable regions}\label{sec:regions}

In this section we discuss protocols operating on layer 4 of our quantum network stack and use GHZ network states to connect routers into regions. Recall that the purpose of this layer is to enable for inter-network graph state requests by LOCC.

We start by proposing a routing protocol for the adaptive phase in quantum networks for quantum routers in Sec. \ref{sec:regions:routing}, which may connect in a highly irregular manner. Next we introduce a way of reducing the size of network states appearing in the static phase for connecting routers in a region in Sec. \ref{sec:regions:hierarchical}, which also reduces the complexity of regions. Finally, we discuss how to achieve reliability for connecting routers in regions in Sec. \ref{sec:regions:reliable}.

\subsection{Region routing}\label{sec:regions:routing}

We start with a brief review of classical routing protocols. In classical routing, there exist protocols using metrics (like e.g. RIP \cite{RIP}) and so-called link state protocols (like e.g. OSPF \cite{OSPF}) - see Sec. \ref{sec:back:classical}.

Protocols using metrics internally construct a so-called routing table. Each table entry is a key-value pair with key corresponding to a network address and values corresponding to distance and interface port of the router to which packets shall be forwarded.

Link-state protocols operate in a different manner. They internally construct a global view of the network topology, i.e. a weighted graph where the weight of an edge corresponds to the distance or cost between two nodes of the network. Depending on the destination IP address of an incoming packet, routers compute a minimal cost path through the network by using Dijkstra's algorithm, see Sec. \ref{sec:back:algorithms}. The protocol which we propose follows a similar approach as link-state protocols in classical networks.

Before we start with the protocol description we first recall that we abstract quantum networks via routers as in classical networks. The router provides, according to our stack of Sec. \ref{sec:stack}, an entry point to a quantum network. For simplicity we assume that there is only one router in each network. Several routers in one network can be taken into account as follows: If there are two routers, then there exists at least one part of the network state which connects those routers. One of both routers can teleport all of its qubit which belong to another region to the other, thereby providing a single entry point to the network under consideration.

\begin{figure}[h!]
\scalebox{1}{
\includegraphics{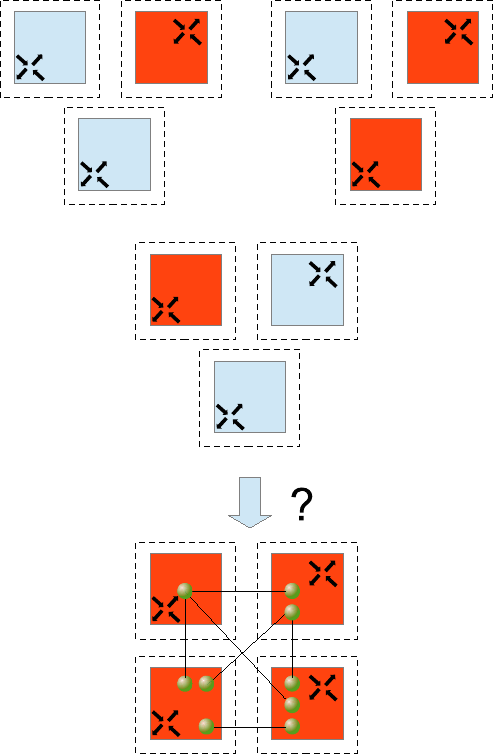}
}
%\flushleft
\caption[h!]{The goal of a quantum routing protocol: A routing protocol shall generate a "virtual network state" between routers of different quantum networks which are  involved in a request. After that the routers combine the virtual network state with the inner network state to transfer the entanglement to the requesting devices. This enables the networking devices to directly apply the graph state linking protocol of Layer 3 to complete the request.\label{fig:routing:goal}}
\end{figure}

In Sec. \ref{sec:stack:layer4} we identified the goal of the network layer, and therefore also routing, as follows: Routing protocols in quantum networks should establish a virtual network state across routers, as this enables routers of networks to combine the virtual network state with the respective inner network states of each router to fulfill graph state requests across network boundaries. The situation is summarized in Fig. \ref{fig:routing:goal}.

Now we discuss a routing protocol which achieves the aforementioned goal. For that purpose we assume that routers of networks connect via GHZ network states in regions. Such a scenario corresponds to the case when a network administrator defines which routers shall connect in a region.

Such a configuration of routers in regions, and thereby also the configuration of network states, may be highly irregular. We stress that such a scenario is of high relevance for practical settings, as it enables network administrators to define network boundaries and which networks connect to each other in regions. The network administrator may have knowledge about the traffic which clients in quantum networks produce and tries to minimize the overall entanglement cost associated with network states.

\begin{figure}[h!]
\scalebox{2.7}{
\includegraphics{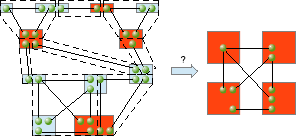}
}
%\flushleft
\caption[h!]{Regions and how they connect: A region (indicated by dashed lines) connects routers (boxes) via a GHZ network state. A router may be part of several regions. If clients of specific networks request a graph state, the region routing protocol shall establish a virtual network state among the routers of these networks. This state can then be used to generate a network state between graph state requesting quantum network devices. \label{fig:routing:ghz:overview}}
\end{figure}

The goal of the routing protocol is now as follows: Clients, possibly located in different networks, wish to generate a particular graph state. The aim of the region routing protocol is to establish a virtual network state across the routers of networks involved in a request, see Fig. \ref{fig:routing:ghz:overview}.

In order to generate such a virtual network state we make use of Dijkstra's algorithm (which also some works on routing for quantum repeater networks use, e.g. \cite{Meter2013b}) and the algorithm for determining a Steiner tree as discussed in Sec. \ref{sec:back:algorithms}. We remark that Steiner trees have also been used in \cite{Bauml17} for deriving fundamental limitations on quantum broadcast channels.  These algorithms require the definition of a cost function $C$ for the edges of the graph. For illustration purposes, we use the number of states the routing protocol consumes as cost, i.e. each edge in the graph has unit cost. We discuss more appropriate cost functions later on.

The input to the region routing protocol is a set of networks (or more precisely, the routers of the networks) which connect via regions. Each router corresponds to a vertex in the graph corresponding to the configuration of regions. Suppose a subset $S = \lbrace N_1, \ldots, N_k \rbrace$ of the vertices request a graph state. Then we perform the following algorithm:

\begin{protocol}[h!]
\caption{RegionRouting($S$)}
\label{protocol:dgrp}
\begin{algorithmic}[1]
    \REQUIRE Set of nodes $S \subseteq V$, $V$ set of  networks
    \STATE Transform the graph of vertices (networks) to a classical graph, where qubits belonging to the same network fusion into one vertex. We denote this graph by $G' = (V,E)$. \label{enu:dgrp:1}
    \WHILE{$S \neq \emptyset$}
    	\STATE Select $v \in S$ \label{enu:dgrp:3}
    	\STATE $T = \mathrm{Steiner}(S,v)$
    	\STATE Generate $\ket{\mathrm{GHZ}_{|S|}}$ with root at $v$ according to $T$ \label{enu:dgrp:5}
    	\STATE $S = S \setminus \lbrace v \rbrace$
    \ENDWHILE
\end{algorithmic}
\end{protocol}

In step \ref{enu:dgrp:1} of protocol \ref{protocol:dgrp} the routers transform the configuration of states connecting the regions to a classical graph by merging all qubits of a router to a single node. However, observe that the routers have to keep track of the states (which correspond to edges in the graph) the Steiner tree algorithm selects internally. The routers further optimize the consumption of states by minimizing the multipartite entanglement they select between regions for a request.
An alternative approach for step \ref{enu:dgrp:1} is to generate the classical graph which the Steiner tree algorithm requires by associating a qubit with a vertex in the classical graph, and creating an edge for every possible Bell-measurement which can, in principle, connect two regions.

Observe that the while-loop of protocol \ref{protocol:dgrp} creates one of the $k-1$ GHZ states of the virtual network state. The steps \ref{enu:dgrp:3} - \ref{enu:dgrp:5} create one part of the GHZ network state with root located at $v'$ by using a Steiner tree between the remaining routers in $S$. A complete example for the routing protocol is provided in Appendix \ref{app:routing}.

We remark that protocol \ref{protocol:steiner} of Sec. \ref{sec:back:algorithms} to determine a Steiner tree requires the usage of a specific cost function. The cost function should take into account for the cost to generate and combine bi- or multipartite states at different layers. Channel noise as well as noise in local operations are relevant for the performance of entanglement distillation protocols and the combination of different states (e.g. via entanglement swapping or merging), and hence determine the cost. We leave a discussion for proper choice of cost function to future work, but remark that the usage of multipartite entangled states can also be beneficial in this respect \cite{Du03,Du05,Duer07}.

\subsection{Hierarchical regions}\label{sec:regions:hierarchical}

One obvious way of connecting quantum networks in a region is to connect all in the same fashion as a quantum network connects its devices, i.e. in a single region. In that case, only $Z$ measurements on the GHZ states connecting the networks are necessary to establish the state depicted in Fig. \ref{fig:routing:goal}. However, such an approach has one serious drawback: The size of the network state will increase with the number of networks, i.e. the number of routers. In particular, to connect $n$ routers in a single region, the largest GHZ state connecting them is of size $n$. In a practical realization, the size of GHZ states might however be limited. The reason is that GHZ states suffer from noise and decoherence, and are in fact more fragile with increasing size $n$ \cite{DB04,He05}.
%Due to the fragility of GHZ states, following such an approach will fail in a noisy, realistic setting \cite{DB04,He05}.
Also the approach discussed in the previous section might be impractical. The network administrator plays a key role, as he defines which routers shall connect in a region. This determines the topology of the network, and a proper knowledge of the underlying traffic is crucial for an efficient choice. However, if the the administrator does not have this knowledge prior to region design, or there are unexpected fluctuations, the topology might be inefficient.

%Another approach of how to connect routers and networks in regions has been discussed in the previous section: A network administrator may want to define which routers shall connect in a region, and which do not. If the administrator has a good knowledge about which networks will request graph states frequently between each others, following such a method will provide an accurate region topology. However, if the administrator does not have this knowledge prior to region design, the topology will be inefficient.

%So we summarize that connecting all routers into one region is inefficient because of ever growing GHZ states, and that a region topology configured by an administrator may also be highly impractical. Therefore, we have to find a solution to solve both of these issues.

The method we propose here is an automatic and efficient scheme for connecting routers into regions in a hierarchical manner. The key element is to use only GHZ states of limited size, and arrange the regions in a hierarchical manner. This avoids the problem of fragile, large GHZ states. In addition, regions can be arranged on demand, e.g. optimized w.r.t. expected traffic. 
We remark that such a hierarchic arrangement was also implicitly assumed in \cite{Wallnofer16_2D,Pirker18}. The features of such hierarchical graphs and their properties in a network structure has recently been analyzed in detail in \cite{Bapat18}.

\begin{figure}[h!]
\scalebox{2}{
\includegraphics{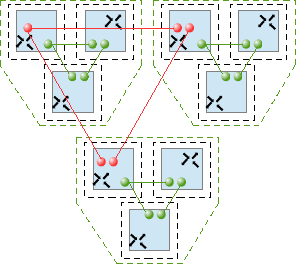}
}
%\flushleft
\caption[h!]{The figure depicts the state for connecting $9$ networks in hierarchical regions for $m=3$: The entire network of routers is broken down into smaller regions, sharing again a network state, with size of at most $3$ (green). One of each routers of a regions connects to the next hierarchical level again via a GHZ network state (red).  \label{fig:routing:ghz}}
\end{figure}

%Such a hierarchical approach for various different graphs appearing at each level, and its evaluation w.r.t. to several graph-related values was analyzed in detail in \cite{Bapat18}.
In our case, regions connect via a GHZ network state, and we fix the maximum number of routers $m$ which are part of such a region. This effectively limits the size of the GHZ network state connecting the routers. Regarding fragility, we remark that in fact a three qubit GHZ state can accept more local depolarizing noise per particle than a Bell pair - only for larger particle numbers there is an increased fragility \cite{Wallnofer16_2D}. The situation for the case of $m=3$ is depicted in Fig. \ref{fig:routing:ghz}.

We can also view these hierarchical regions as substructures inside the network layer of the quantum networking stack. More specifically, the hierarchical layers may be considered as high layers in the stack.
We call a router which connects regions at different hierarchical levels also a designated router (DR). These routers enable to reach routers/networks which are located at regions at different levels of the hierarchy.

The process of establishing and connecting routers in regions can be done automatically: If a new routers starts, it simply discovers all previously appeared routers classically. Then, the designated routers checks if the new router fits into any of the existing regions. If so, then the new router will be added to that region. If not, a new hierarchical level is created, thereby creating an $m-$ary tree of regions.
%This procedure is now applied iteratively. Regions connect via GHZ network states of upper bounded size of at most $m$. If, at the lowest level, a new router does not fit into any region, a new region at the lowest level is created and hierarchically connected, thereby creating an $m-$ary tree of regions. 
Notice that in principle, one can choose the position of routers in the hierarchy also w.r.t. to certain parameters, like e.g. traffic. This enables one to place routers of networks which often demand requests in the same region, thereby optimizing the regions between quantum networks. We also observe that the complexity of the region routing protocol decreases due to such a hierarchy of regions.

\subsection{Reliable regions}\label{sec:regions:reliable}

Finally we want to discuss the reliability of regions. In principle, since regions connect via GHZ network states, they will suffer from similar problems as quantum networks at layer 3 if a router disconnects.

The schemes we discussed in Sec. \ref{sec:layer3:reliable} enable for compensating the fail of network devices in a network, leaving the remaining devices within the network with a functional network state.

However, in the setting of regions the situation is more involved, since the failure of one router at the boundary of a region will disable also all other routers of the same region to generate graph states to other regions.

\begin{figure}[h!]
\scalebox{1.3}{
\includegraphics{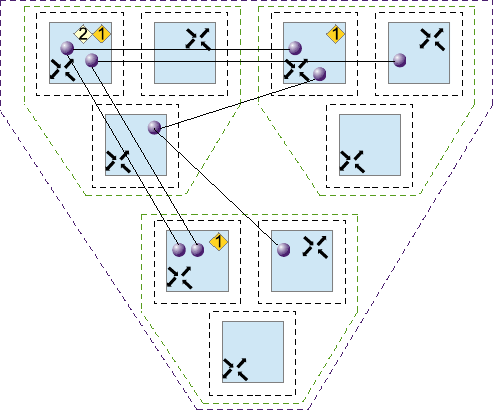}
}
%\flushleft
\caption[h!]{Symmetrization of a network state between regions (shown for three qubit GHZ state): Symmetrization is done w.r.t. all possible permutations of routers residing in different regions. Observe that the inter-region entanglement is still of GHZ type. \label{fig:ghzregion:sym}}
\end{figure}

Such an issue can be solved via symmetrization inside regions. In particular, suppose $m$ regions connect with a GHZ network state. Instead of distributing the qubits of the GHZ network states to one particular router of each region, we symmetrize the GHZ states inside the respective regions. For example, in the case of the regions $A, \ldots, M$ we distribute the qubits of the largest GHZ to different routers inside the regions $A, \ldots, M$. In particular, for region $A$, we can assign the qubits to $|A|$ different routers, in region $B$, we can distribute the qubits to $|B|$ different routers and so on. The scheme is illustrated in Fig. \ref{fig:ghzregion:sym}.

This result in total in
\begin{align}
|A| \cdot |B| \cdot \ldots \cdot |M|
\end{align}
different $M$ qubit GHZ states. Observe that the GHZ-type of entanglement between the regions $A, \ldots, M$ is preserved by this symmetrization. The same procedure is also applied to the smaller GHZ states between regions.

This technique will introduce complexity to the region routing protocol and therefore, one may use only one specific permutation in regular operation mode, e.g. the configuration $A_1, \ldots , M_1$. All other permutations are only used in case of route failures. 

The approach which we presented in this section shall be considered as a proof-of-principle for achieving reliability between regions, and other approaches may yield better results. For example, one may consider to apply  quantum error correction codes for the erasure channel \cite{Grassl97} in the context of reliable regions. In such a scenario, each qubit of the GHZ state between regions could be encoded in a logical qubit of the quantum error correction code for the erasure channel, and the physical qubits comprising the logical qubit be distributed among the routers inside the region. This allows to compensate for failures of router inside regions. However, in order to recover from a failure, the remaining routers within a region have to communicate and collaborate to restore the GHZ state between regions, which will necessarily consume entanglement inside a region. Finally, we emphasize that we do not claim optimality of the symmetrization approach of this section to achieve reliability between regions. 

%However, note that multiple copies of the network states between regions are required anyway, so this protocol does not introduce an additional cost or effort.

\section{Auxiliary -- Reachability protocols \label{sec:aux:reachability}}

In this section we discuss protocols for checking whether entanglement connectivity between quantum networking devices is still present or not.
Such a quantum ping is of high importance for several layers, which motivates us to put them as auxiliary protocols in our proposed quantum stack. In particular, at layer 2 of the quantum networking stack one may use quantum repeaters to establish entanglement between remote nodes. At layer 3, the switches of a network might also want to check whether the entanglement among each other is still present. Finally, routers could want to determine whether a network is reachable via some intermediate regions. For that purpose, first routing towards a network has to be done.

In classical networks, the reachability of network devices is achieved by so-called pings \cite{icmp}. The ping mechanism is part of the Internet Control Messaging Protocol \cite{icmp} which operates above the Internet Protocol \cite{ip} at layer 3 of the classical network stack. The protocol is rather simple: If a network device or an operator wants to check whether a destination is reachable, it sends an echo-request to the corresponding destination IP address. The networking devices determine a path to the destination by routing the packet. Once the echo-request reaches the destination, it replies to the echo-request with an echo-reply. The reply is routed back to the originator of the request, and therefore connectivity/reachability has been checked.

For quantum networks the situation is more involved, as the network relies on entanglement. In particular, a quantum ping protocol shall ensure not only the classical reachability of a device, but also that entanglement between networking devices is still present and usable for future requests. In general we identify three different levels of checking for reachability:

\begin{enumerate}
	\item Classical ping: This corresponds to the classical ICMP echo protocol.
	\item State verification: This level has the purpose to check whether entanglement is still available or not, where the local apparatus is trusted. Routers can achieve this by using techniques like entanglement witnesses, checking the correlation operators of graph states, or quantum state tomography.
	\item Device-independent state verification: This level aims at checking for entanglement without trusting the local apparatus of each network device, and is therefore the strongest level of reachability.
\end{enumerate}

We use these levels to illustrate the concept of reachability in quantum networks. A detailed analysis of these levels we leave for future work. Nevertheless, we outline the basic ideas and concepts which led to this categorization here.

We remark that in general a non-destructive state verification without additional entanglement as resource is not possible. More specifically, without using entanglement from the network, it is in general not possible to reliably guarantee that entanglement between the networking devices is still present. We can only locally measure (and hence destroy) a subset of the quantum states. Under the assumption the initial ensemble consists of independent copies, one can then make predictions about the state and quality of the remaining ensemble. The number of states which such a technique consumes highly depends on the protocol the networks use. However, if the network devices are free to consume some of the entanglement in the network (or possibly across different networks), non-destructive state verification is feasible. If packets, like e.g. measurement outcomes or acknowledgements, get lost during the reachability protocol runs, the protocol may have to restart. We emphasize that such a decision heavily depends on the ping protocol, and when considering an implementation of a quantum ping, one shall consider reliable classical transport protocols like TCP \cite{tcp}.

For graph states, methods for state verification have been put forward, see e.g. \cite{McKague10,McKague16,Markham18,Pal14}. These approaches are based on stabilizer measurements and their expected values, or tests of generalized Bell inequalities. The latter allows for device independent state verification.

\section{Conclusion}\label{sec:conclusion}

In this work we have introduced a stack model for quantum networks, which is especially well suited for networks relying on multipartite entanglement. The stack comprises several different, goal-oriented layers. Each of these layers has a clear responsibility, and therefore enables for a straightforward implementation. By introducing such a layered model, we have broken down the complexity of large-scale quantum networks into smaller, manageable pieces, where each layer can be studied independently. Such a model exploits the power of abstracting problems, in which high-level layers do not have to deal with concrete implementations of low-level protocols. Furthermore, due to concentrating physical implementation details relating to interfacing with quantum channels, we have obtained an abstract stack model of a quantum network in terms of layers where high-levels layers are not concerned with details of how to access quantum channels or to convert between memory and transmission technologies.  

Furthermore we have introduced several protocols at different layers. In particular, we have found a way for ensuring reliability inside and between different networks. We also defined the term routing between quantum networks connecting via multipartite quantum states. To solve this routing problem we presented a protocol which achieves the goal by using Steiner tree constructions.

It remains to be considered which cost function is appropriate for the Steiner tree construction. Furthermore, it might be interesting to find alternative multipartite quantum states for connecting devices in a network, and routers in regions, as well as identifying other approaches for ensuring reliability in quantum networks.

\section*{Acknowledgements}
This work was supported by the Austrian Science Fund (FWF): P28000-N27 and P30937-N27. We thank J. Walln\"ofer for helpful discussions and comments on the manuscript.

\appendix

\section{An example of region routing}\label{app:routing}

\begin{figure}[h!]
\scalebox{1.7}{
\includegraphics{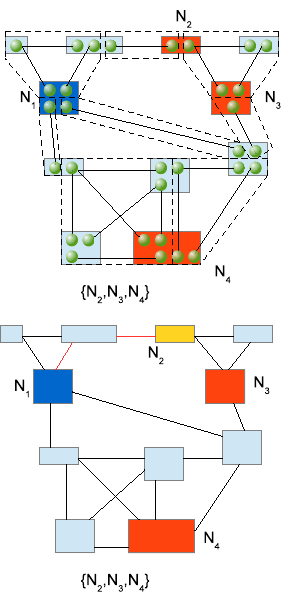}
}
%\flushleft
\caption[h!]{Clients of the networks $N_1, \ldots, N_4$ want to generate a graph state. The configuration of routers (boxes) in regions (dashed lines) is shown at the top of the figure. There are four regions of size two, three regions of size three (one includes $N_1$, the other one includes $N_2$ and $N_3$) and one of size four. Observe that networks may be part of several regions, like e.g. $N_1$, which connect at one router, see the discussion above. On the bottom we depict the first step of the protocol: The protocol starts with $S=\lbrace N_1, \ldots, N_4 \rbrace$, and selects $N_1$ to be the root of the first Steiner tree.  Then, in the first step of the Steiner tree computation, $N_2$ is chosen to be closest network to $N_1$, and the path is added to the Steiner tree (red edge). \label{fig:routing:ghz:steiner:1}}
\end{figure}

We illustrate the region routing protocol  \ref{protocol:dgrp} for a concrete example, which is shown in Fig. \ref{fig:routing:ghz:steiner:1}. In this example, four clients of four different networks want to share a graph state. The region routing protocol shall establish a virtual network state among the respective routers. The first step in the Steiner tree construction for the first router is shown in Fig. \ref{fig:routing:ghz:steiner:1}.

%\begin{figure}[h!]
%\scalebox{0.7}{
%\includegraphics{dgrp_steiner_2.eps}
%}
%%\flushleft
%\caption[h!]{The figure depicts the remaining algorithm to finalize the Steiner tree among all four networks. Iteratively, one vertex is added to the Steiner tree. \label{fig:routing:ghz:steiner:2}}
%\end{figure}

After the first Steiner tree construction finishes, the algorithm transforms the Steiner tree to a GHZ state as follows: If the degree of the root vertex $r$ selected by the protocol is larger than one, it combines all qubits into one (by e.g. preparing a local GHZ state of size $\deg(r) +1$, and performing at most $r$ Bell-measurements), thereby again obtaining a GHZ state of size $\deg(r) + 1$. All nodes which are not target networks, employ Bell-measurements to their qubits which have been selected for the Steiner tree. If a target node $t$ is not a terminal node, then it locally creates a GHZ state of size $\deg(t) +1$, keeps one qubit, and performs at most $t$ Bell-measurement of all qubits which are part of the Steiner. Observe that it may be necessary that some routers measure qubits in the $Z$ basis to shape the GHZ states in regions.

%\begin{figure}[h!]
%\scalebox{1.5}{
%\includegraphics{dgrp_steiner_first.eps}
%}
%%\flushleft
%\caption[h!]{Generating the first part the GHZ network state $\ket{\mathrm{GHZ}_4}$: The router of $N_2$ has to create a local three qubit GHZ state, keeps one qubit, and performs a Bell-measurement on all remaining ones. The yellow cloud corresponds to a measurement in the $Z$ basis. All other qubits along the path get combined via Bell-measurements. The result is a four qubit GHZ state between all networks.\label{fig:routing:ghz:steiner:firstdone}}
%\end{figure}

\begin{figure*}[htpb]
\scalebox{2}{
\includegraphics{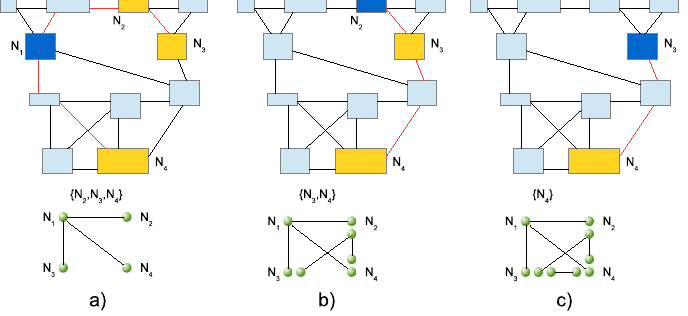}
}
%\flushleft
\caption[h!]{The Steiner tree constructions for all GHZ network states of our example.In a) the state $\ket{\mathrm{GHZ}_4}$, in b) the state $\ket{\mathrm{GHZ}_3}$ and c) a Bell-pair is generated. After the protocol finishes, the router share a full GHZ network state. \label{fig:routing:ghz:steiner:restdone}}
\end{figure*}

In the next step, the protocol \ref{protocol:dgrp} removes $N_1$ from the set of target networks $S$ and we perform the same procedure to the remaining network nodes in $S$. The final state after the routing protocol is depicted in Fig. \ref{fig:routing:ghz:steiner:restdone}.

\section{How things combine: A full example of how the quantum network stack works}\label{app:stack}

In this section we provide a full example of how the quantum network stack presented in this paper works. For that purpose we illustrate the responsibilities of each layer and the corresponding operations of the network devices.

\begin{figure}[h!]
\scalebox{2.2}{
\includegraphics{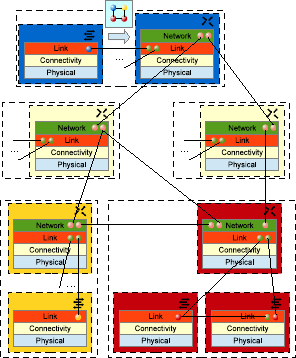}
}
%\flushleft
\caption[h!]{The figure depicts the example which we use to illustrate how the quantum network stack works. In this example, a client of the switch in the blue network requests to share a four qubit cluster state with a client located at a switch in the yellow, and two clients located at two different switches in the red network. \label{fig:ex:1}}
\end{figure}

We start with the example scenario depicted in Fig. \ref{fig:ex:1}.

As explained in the figure, a client located in the blue network wants to share a four qubit cluster state with clients located in the yellow and red network. The routers connect in regions via GHZ network states in an irregular manner, defined by a network  administrator. Observe that the blue network connects to the red and yellow network via some intermediate networks.

\begin{figure}[h!]
\scalebox{2.2}{
\includegraphics{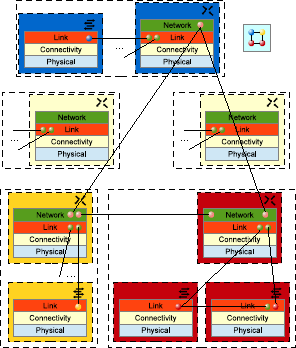}
}
%\flushleft
\caption[h!]{The situation after routing completes: The routers of the blue, yellow and red networks share a virtual network state. \label{fig:ex:2}}
\end{figure}

The first step in processing the request is now to run the routing protocol of layer 4. The result of routing is a virtual network state between the routers of the blue, yellow and red network, see Fig. \ref{fig:ex:2}.

\begin{figure}[h!]
\scalebox{2.2}{
\includegraphics{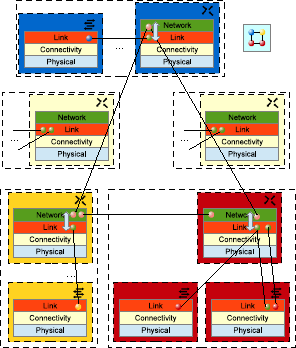}
}
%\flushleft
\caption[h!]{How the routers combine the virtual network state and the respective local, intra-network states (we illustrate the principle for the largest GHZ state of the virtual network state): The double arrows indicate between which qubits the routers perform a Bell-measurement. Observe that such merging operations might also induce the creation of some local GHZ states for expanding states at the network boundary. However, the procedure will always result in a GHZ state spanning all network devices involved in the request. \label{fig:ex:3}}
\end{figure}

After routing completes, the routers fulfill their second responsibility: They merge the virtual network state with their respective intra-network states to transfer the entanglement of the virtual network state to the requesting network devices inside their networks. This is done by performing Bell-measurements between the qubits of the network layer and the qubits of the link layer, which is depicted in Fig. \ref{fig:ex:3}.

\begin{figure}[h!]
\scalebox{2.2}{
\includegraphics{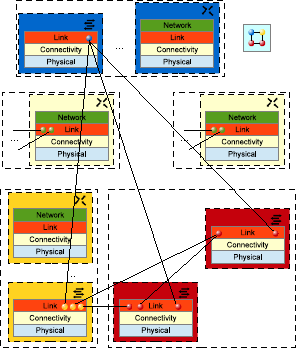}
}
%\flushleft
\caption[h!]{Once the routers have combined the entanglement of the virtual network state among the routers with the network states of their associated networks, the switches of different networks share a GHZ network state. Observe that we have depicted in this figure a full GHZ network state for illustration. In real world applications, only a subset of these states might be necessary to complete the graph state request. \label{fig:ex:4}}
\end{figure}

Once this operations complete, the requesting switches share one copy of the GHZ network state among each other. This operation is now carried out several times to generate multiple copies of the GHZ states, where the number of iterations depends on the graph state request. Observe that such a full GHZ network state enables the devices to generate any arbitrary graph state among the requesting network devices by simply invoking the linking protocol of layer 3 (see \cite{Pirker18} or Sec. \ref{sec:stack:layer3}). The situation is shown in Fig. \ref{fig:ex:4}.

\begin{figure}[h!]
\scalebox{2.2}{
\includegraphics{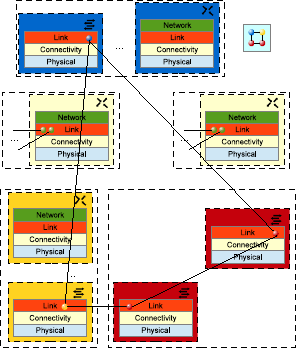}
}
%\flushleft
\caption[h!]{The switches have completed the graph state request by generating the four qubit cluster state via the state linking protocol of layer 3. For that purpose they consume the network state as established in Fig. \ref{fig:ex:4}. Finally, the network devices teleport this state to the requesting clients. \label{fig:ex:5}}
\end{figure}

Finally, as outlined above, the switches invoke the state linking protocol of layer 3 (i.e. the graph state generating protocol of the layer they are operating on) to complete the graph state request for the requesting clients. The final graph state is shown in Fig. \ref{fig:ex:5}.

After the graph state request finishes, entanglement has to be replenished in most cases (dynamic phase). For that purpose, the switches and routers which were involved in the request instruct the layer 2 devices, i.e. the quantum repeaters, to re-establish the required entanglement in the networks and regions. In doing so, the quantum network devices use the physical channel configuration of the network at layer 1 to send quantum states to neighbouring devices. For example, in the blue network it may suffice to re-establish Bell-pairs between the router and the switch. In contrast, in the red network the networking devices instruct the repeaters of the network to generate the required resource mandatory to generate the consumed GHZ state among these devices. Finally, the routers instruct the repeaters residing in regions to generate the entanglement which is required to re-establish the network states connecting routers of a region.

\bibliographystyle{apsrev4-1}
\bibliography{quantum_networks}

\end{document}